\documentclass[sigconf,authorversion]{acmart}
\usepackage{algorithmic,algorithm}
\usepackage{caption}
\usepackage{tabularx}
\usepackage{graphics}
\usepackage{multirow}

\AtBeginDocument{%
  \providecommand\BibTeX{{%
    \normalfont B\kern-0.5em{\scshape i\kern-0.25em b}\kern-0.8em\TeX}}}

\copyrightyear{2021}
\acmYear{2021}
\setcopyright{acmlicensed}\acmConference[CHI '21]{CHI Conference on Human Factors in Computing Systems}{May 8--13, 2021}{Yokohama, Japan}
\acmBooktitle{CHI Conference on Human Factors in Computing Systems (CHI '21), May 8--13, 2021, Yokohama, Japan}
\acmPrice{15.00}
\acmDOI{10.1145/3411764.3445569}
\acmISBN{978-1-4503-8096-6/21/05}

\begin{document}

\title{Designing Effective Interview Chatbots: \\ Automatic Chatbot Profiling and Design Suggestion Generation for Chatbot Debugging}

\author{Xu Han}
\affiliation{%
  \institution{University of Colorado Boulder}
  \city{Boulder}
  \state{CO}
  \country{USA}}
\email{xuha2442@colorado.edu}

\author{Michelle Zhou}
\affiliation{%
  \institution{Juji, Inc.}
  \city{San Jose}
  \state{CA}
  \country{USA}}
\email{mzhou@acm.org}

\author{Matthew J. Turner}
\affiliation{%
  \institution{University of Colorado Boulder}
  \city{Boulder}
  \state{CO}
  \country{USA}}
\email{matthew.turner@colorado.edu}

\author{Tom Yeh}
\affiliation{%
  \institution{University of Colorado Boulder}
  \city{Boulder}
  \state{CO}
  \country{USA}}
\email{tom.yeh@colorado.edu}



\begin{abstract}

Recent studies show the effectiveness of interview chatbots for
information elicitation. However, designing an effective interview
chatbot is non-trivial. Few tools exist to help designers 
design, evaluate, and improve an interview chatbot iteratively. Based on a
formative study and literature reviews, we propose a computational
framework for quantifying the performance of interview
chatbots. Incorporating the framework, we have developed
\textit{iChatProfile}, an assistive chatbot design tool that can
automatically generate a profile of an interview chatbot with
quantified performance metrics and offer design suggestions for improving
the chatbot based on such metrics. To validate the effectiveness of \textit{iChatProfile}, we
designed and conducted a between-subject study that compared the
performance of 10 interview chatbots designed with or without using
\textit{iChatProfile}. Based on the live chats between the 10 chatbots
and 1349 users, our results show that \textit{iChatProfile} helped the
designers build significantly more effective interview chatbots, improving
both interview quality and user experience.
\end{abstract}



\begin{teaserfigure}
     \centering
     \begin{minipage}[b]{0.45\textwidth}
         \centering
         \includegraphics[width=1.1\textwidth]{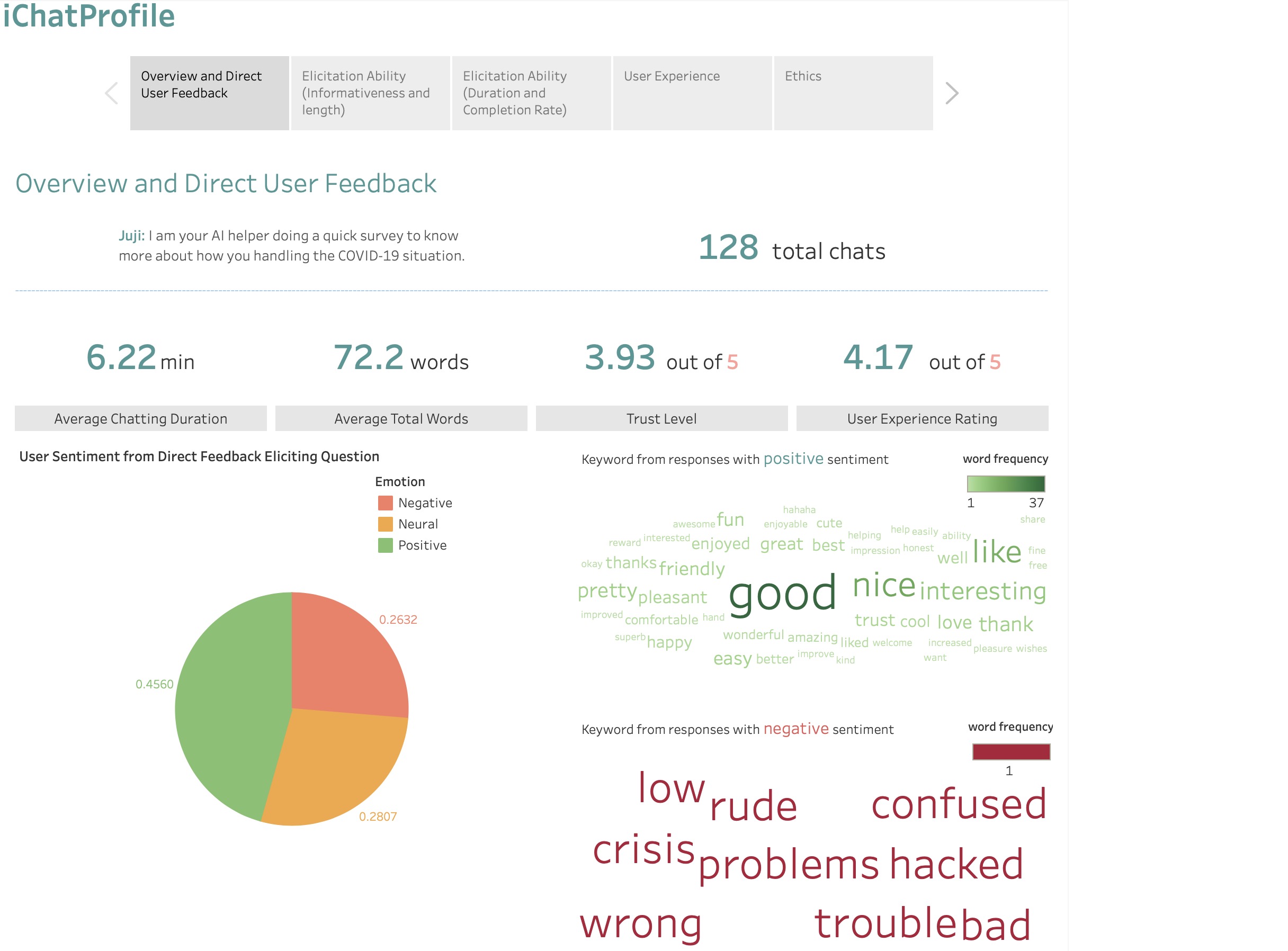}
         \caption*{(a)}
         \Description{The example profile is mainly composed of 4 parts, including a brief introduction of the chatbot overview on the top, a summarized chatbot pilot results in the middle , a pie chart of interviewee sentiments and word clouds at the bottom.}
     \end{minipage}
     \hspace{0.025\textwidth}
     \begin{minipage}[b]{0.45\textwidth}
         \centering
         \includegraphics[width=1.1\textwidth]{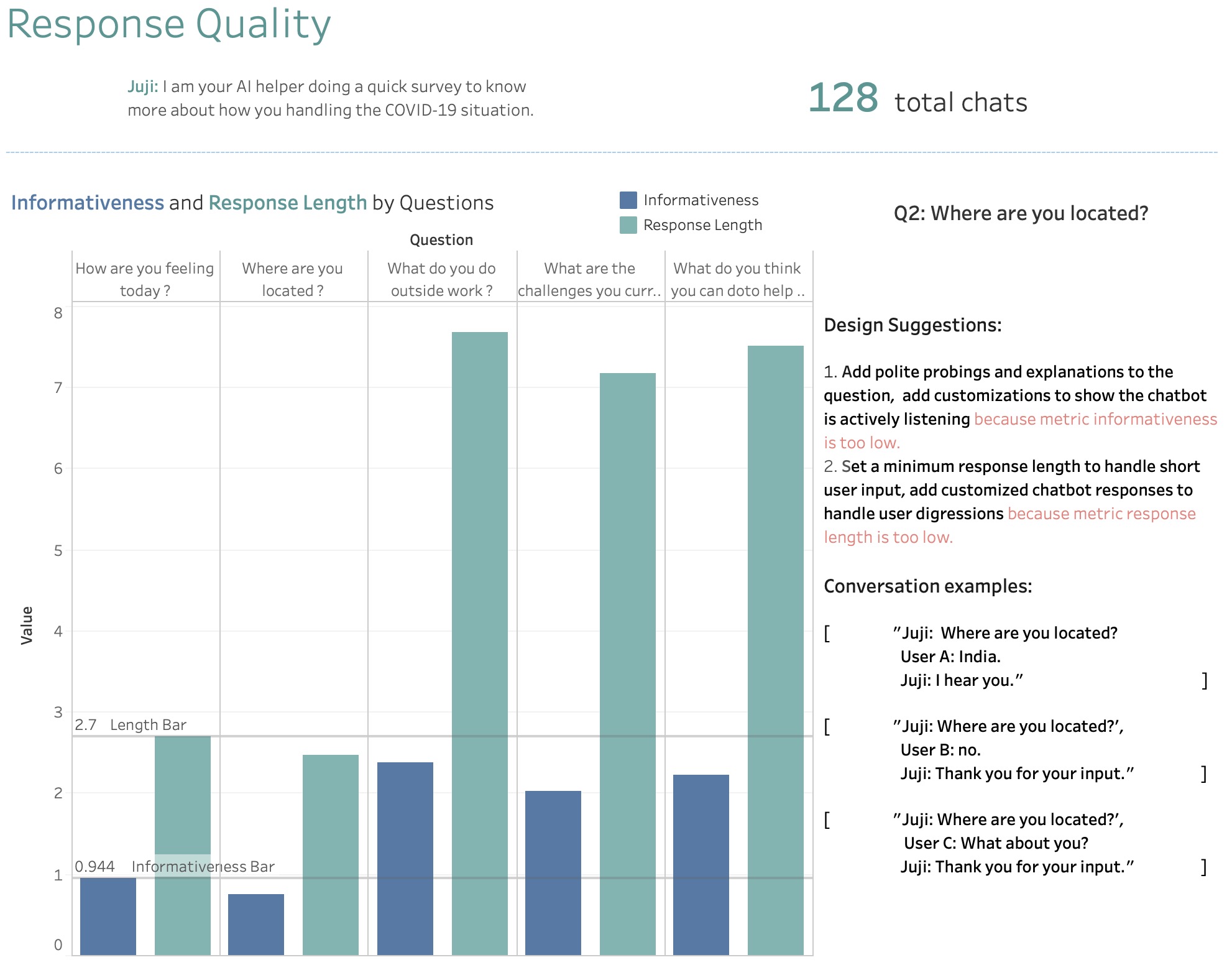}
         \caption*{(b)}
         \Description{The example contains mainly two parts. The left is a bar chart indicating the response quality while the right part is the generated design suggestions.}
     \end{minipage}
     \caption{\textit{iChatProfile} dashboard. (a) an example chatbot profile. (b) examples of auto-generated design suggestions for improving a chatbot.}
     \label{fig:3}
\end{teaserfigure}

\maketitle

\section{Introduction}

During the past few years, chatbots have been used to conduct
interviews by engaging users in one-on-one text-based
conversations \cite{Xiao2019-hk}. Recent studies show that 
interview chatbots are more effective at engaging users and eliciting quality information
from the users, compared to traditional online surveys
\cite{Xiao2019-hk, Kim2019-rs, Xiao2019-ff}. 

Despite their promises, it is challenging and time consuming to build
effective interview chatbots due to the limitations in today's
technologies and the complexity involved in interview conversations
\cite{Xiao2020-fr, Grudin2019-dx}. Like building any complex
interactive systems \cite{nielsen1993iterative}, one potential
approach is to design and improve an interview chatbot iteratively. 
Specifically, the iterative design of an interview chatbot is to fulfill two main goals. First, like designing any
user interviews or surveys \cite{chen2011finding, nielsen1993iterative}, designers of an interview chatbot need to
ensure the effective design of an interview task (e.g., proper and clear
wording of questions). Second, like building any conversational agents \cite{jurafsky2017dialog}, designers of an interview chatbot need to make sure that the chatbot can successfully carry out such an interview task \cite{Xiao2020-fr}. 

\begin{figure*}
     \centering
     \begin{minipage}[b]{0.3\textwidth}
         \centering
          \includegraphics[width=\textwidth]{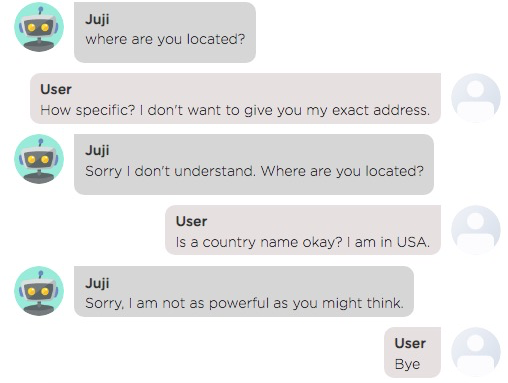}
         \caption*{(a)}
         \Description{A screenshot of Juji chatting dashboard with poorly handled conversations.}
         
     \end{minipage}
     \hfill
     \begin{minipage}[b]{0.3\textwidth}
         \centering
         \includegraphics[width=\textwidth]{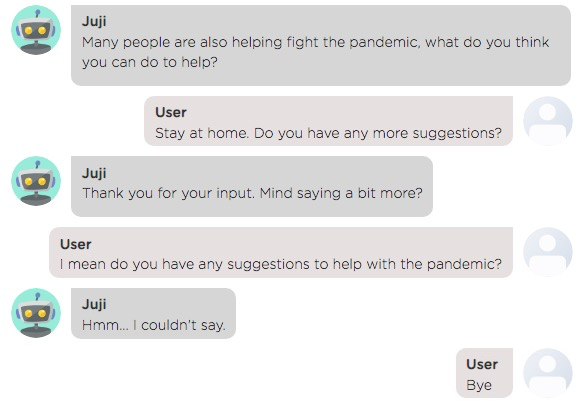}
         \caption*{(b)}
         \Description{A screenshot of Juji chatting dashboard with poorly handled conversations.}
     \end{minipage}
     \hfill
     \begin{minipage}[b]{0.3\textwidth}
         \centering
         \includegraphics[width=0.95\textwidth]{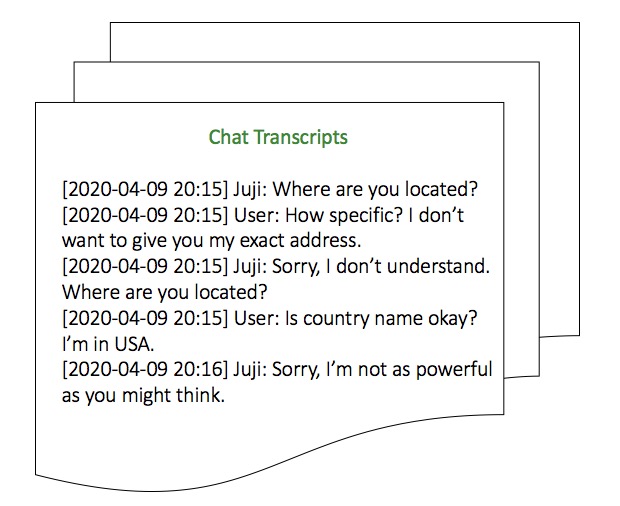}
         \caption*{(c)}
         \Description{An example of raw chat transcripts.}
     \end{minipage}
     \caption{Examples of poorly handled conversations by an interview chatbot. (a) vague question and (b) unhandled user input. (c) chat transcripts set in practice with poorly handled conversations}
     \label{fig:badexample}
\end{figure*}

To achieve above goals, designers often conduct
pilot studies prior to a formal study \cite{presser2004methods}. However, interview chatbot designers face two challenges in detecting
let alone fixing the potential issues revealed by the pilot
studies. First, designers must examine chat transcripts to discover
potential issues and it is laborious and time consuming to do so manually. For example, Fig~\ref{fig:badexample}(a) shows
that an interview question poses a challenge for a user due to a lack of clarity, while Fig~\ref{fig:badexample}(b) shows a chatbot-unrecognized user input during an interview, which could result in
poor user experience or even abandoned interviews. To detect such issues in practice, chatbot designers must examine chat transcripts (Fig~\ref{fig:badexample}(c)) to discover
them. It is laborious and time consuming to do so manually especially if the designers need to detect such issues from a large number of chat transcripts. Second, even if
the designers have discovered such issues from reading chat transcripts, they
might not know how to fix the issues and improve the chatbot due to a lack of relevant experience (e.g., designing effective interview interactions).

To address the above two challenges, we have been developing a tool, called \textit{iChatProfile}, which can aid chatbot designers in building, evaluating, and improving
interview chatbots iteratively.  In particular, \textit{iChatProfile} \textit{automatically}
analyzes chat transcripts and computes a set of chatbot performance metrics to present designers with a chatbot profile (Fig~\ref{fig:3}(a)). Based on the chatbot profile, \textit{iChatProfile} also \textit{automatically} generates a set of design suggestions to guide designers to improve the chatbot (Fig~\ref{fig:3}(b)).

In this paper, we present the key
steps taken to build \textit{iChatProfile}. First,
we present a formative study that was conducted to understand the difficulties
that designers face when building an interview chatbot and identify
their desired design assistance. Second, we describe a computational framework that quantitatively measures the effectiveness of an interview chatbot from multiple dimensions, including elicitation ability, user experience, and ethics.  Third, we present \textit{iChatProfile} that was built based on our formative study and the computational framework. To validate the effectiveness of
\textit{iChatProfile}, we designed and conducted a between-subject
user study that compared the performance of chatbots designed with or
without using \textit{iChatProfile}. A total of 10 chatbots were created and evaluated live by
1349 participants from Amazon Mechanical Turk. We compared the
performance of these chatbots. The results show that the chatbots
designed with the help of \textit{iChatProfile} performed
significantly better along many dimensions, including improved user response quality and user experience.

To the best of our knowledge, our work is the first on building an assistive
design tool for creating interview chatbots. As a result, our
work reported here provides three unique contributions:

\begin{enumerate}
\item{\textit{A computational framework for quantifying the effectiveness of interview chatbots}. This framework
  comprehensively evaluates the effectiveness of an interview chatbot by computing an extensive set of performance metrics covering multiple dimensions: elicitation ability, user experience, and ethics. Other chatbot researchers and practitioners can easily adopt
  or extend this framework to build their own chatbot evaluation tools.}

\item {\textit{Practical approaches to assisting iterative design of
    interview chatbots}. \textit{iChatProfile} presents practical
  implementations of an assistive chatbot design tool. Because we have
  demonstrated the effectiveness of our implementations, others could replicate or extend
  our approaches to create more tools aiding chatbot design.}

\item {\textit{Design implications for building assistive chatbot
  design tools beyond building interview chatbots.} Although our
current work focuses on aiding the design of interview chatbots, it
presents design considerations for assisting the design of 
other types of chatbots, such as counseling or training
chatbots, which share similar design requirements (e.g., ethical considerations).}

\end{enumerate}

\section{Related Work}

\subsection{Chatbots for Information Elicitation}
AI-powered conversational user interfaces, also known as AI chatbots or chatbots for short, allow users to communicate with computers in
natural language, providing more flexible
\cite{brennan1990conversation} and personalized user experience
\cite{Zhang2018-ir}. Such benefits have encouraged the creation of a
wide array of chatbot applications, such as virtual assistants
\cite{Liao2018-jc}, social companions \cite{Shum2018-wt}, and
interview chatbots \cite{Li2018-gd}.  Our work is most relevant to the
use of chatbots for information elicitation \cite{Xiao2019-hk,
  Xiao2019-ff, Kim2019-rs}.

Researchers have developed various chatbots to elicit information from
users through text-based conversations. For example, Bohus and
Rudnicky introduce dialog systems that gather required information for
performing specific tasks (e.g., making travel
reservations) \cite{bohus2009ravenclaw}. More recently, a number of
interview chatbots have been developed to elicit information from a
target audience.  For example, a chatbot is built to interview
students for effective teaming \cite{Xiao2019-ff} and another chatbot
to interview gamers for eliciting their game opinions
\cite{Xiao2019-hk}. Williams et al. have developed a chatbot to
interview employees for workplace productivity
\cite{williams2018supporting}. Compared to traditional, static online
surveys, these interview chatbots enhance information elicitation
\cite{Xiao2019-ff, Kim2019-rs} by providing interactive
feedback \cite{conrad2005interactive} and asking follow-up questions
\cite{oudejans2011using}.

Our work is directly related to the efforts of creating interview
chatbots.  However, existing work focuses on developing
interview chatbots for specific information elicitation tasks (e.g.,
\cite{Xiao2019-ff, Xiao2019-hk, williams2018supporting}) or powering
interview chatbots with specific skills (e.g., giving them a personality \cite{zhou2019trusting} and active listening
skills \cite{Xiao2020-fr}). While we learn from these efforts, our
work reported here has a very different focus: we want to build a tool
that can automatically evaluate the performance of an interview
chatbot and provide design suggestions for improving the chatbot.

\subsection{Chatbot Platforms}
There are a number of chatbot platforms and these platforms can be
broadly divided into three categories. First, chatbot platforms like
Chatfuel \cite{noauthor_chatfuel} and ManyChat
\cite{noauthor_manychat} allow non-IT professionals to
build a chatbot without coding. Since these platforms provide
limited AI/NLP capabilities, it would be difficult to create
interview chatbots that can understand users especially when
open-ended interview questions are involved. The second type includes platforms like Google Dialogflow
\cite{noauthor_dialogflow} and IBM Watson
\cite{noauthor_IBMwaston}. These platforms provide designers with more
flexibility to customize a chatbot's AI/NLP capabilities but designers
must have basic AI/NLP knowledge to use the tools. The third
category includes platforms like Juji, which provides a rich set of pre-built
AI capabilities to enable non-IT designers to build chatbots without
any expertise of AI/NLP \cite {noauthor_Juji}. While chatbot designers
can choose to use any of the chatbot platforms, none of the platforms
provides a tool like ours reported here: a tool that helps designers
evaluate a chatbot's performance and provides design suggestions to
improve the chatbot.

\subsection{Evaluating Conversational AI Systems}
Researchers have developed a number of approaches to evaluating
conversational AI systems. These approaches can be roughly organized
into two categories: measuring objective system performance (e.g.,
task completion rate for task-oriented chatbots
\cite{jurafsky2017dialog}) and assessing subjective human experience
(e.g., measuring users' trust in a chatbot
\cite{zhou2019trusting}). Incorporating multiple metrics, evaluation
frameworks have also been proposed to systematically measure the
performance of conversational AI systems. For example, PARADISE has
been used to evaluate the performance of task-oriented, spoken dialog
systems \cite{walker1997paradise}, typically developed by AI/NLP experts.  
Unlike these works that evaluate
conversational AI systems in general, our work presented here focuses
on evaluating the performance of interview chatbots, typically designed
by non-IT professionals. While we borrow
some of the existing objective and subjective metrics, we have
developed a computational framework specifically for quantifying the
performance of interview chatbots with \textit{actionable
  insights}---design suggestions that can help designers improve an
interview chatbot.

\subsection{Design Suggestion Generation for Effective Interaction}

Our work on generating design suggestions is also related to
various efforts on guiding the design of human-computer
interfaces, such as chatbot systems \cite{han2019evaluating} and
graphical user interfaces (GUI) \cite{Lee2020-mk, Xu2014-fc}. For
example, Han et al. combine domain-specific knowledge together with
observational studies to generate rule-based design suggestions for
task-oriented chatbots \cite{han2019evaluating}. One of the drawbacks
of this approach lies in its inflexibility of adapting design suggestions
to changing design goals or dynamic design issues occurring in real
time. On the other hand, Lee et al. use autoencoder and k-nearest
neighbor algorithms to recommend GUI design examples that help designers
in real time \cite{Lee2020-mk}. Moreover, Xu et al. have developed a
system that incorporates crowdsourcing to generate design suggestions
for GUI designers \cite{Xu2014-fc}. While we learn from these
approaches, we are unaware of any approach to automatic generation
of design suggestions based on computed chatbot performance as our 
approach does.

\begin{figure*}
     \centering
     \begin{minipage}[b]{0.32\textwidth}
         \centering
         \includegraphics[width=\textwidth]{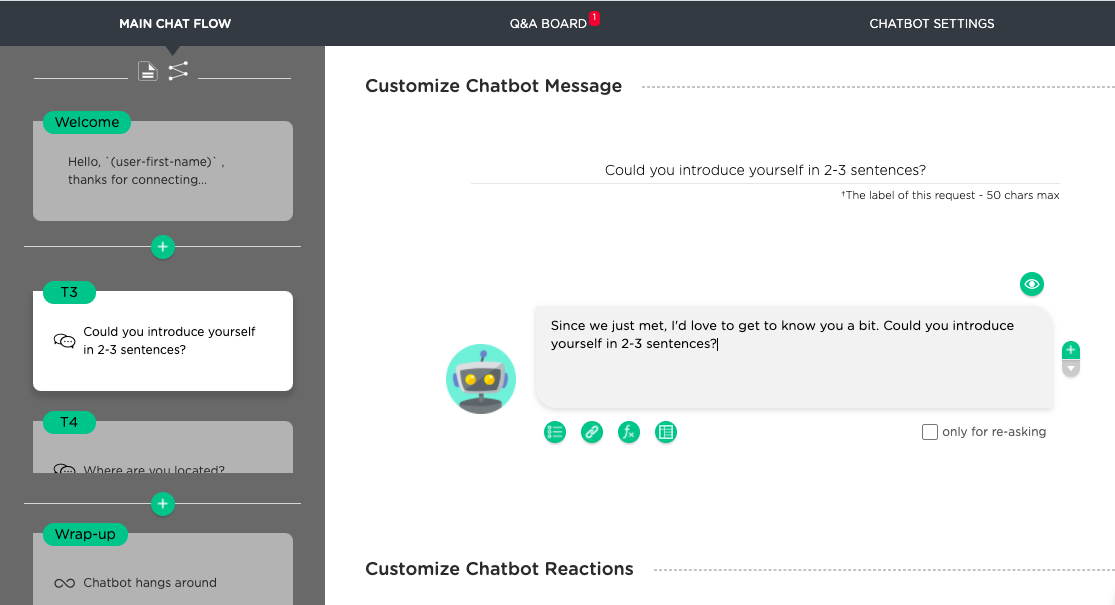}
         \caption*{(a)}
         \Description{A screen shot of Juji dashboard}
     \end{minipage}
     \hfill
     \begin{minipage}[b]{0.32\textwidth}
         \centering
         \includegraphics[width=\textwidth]{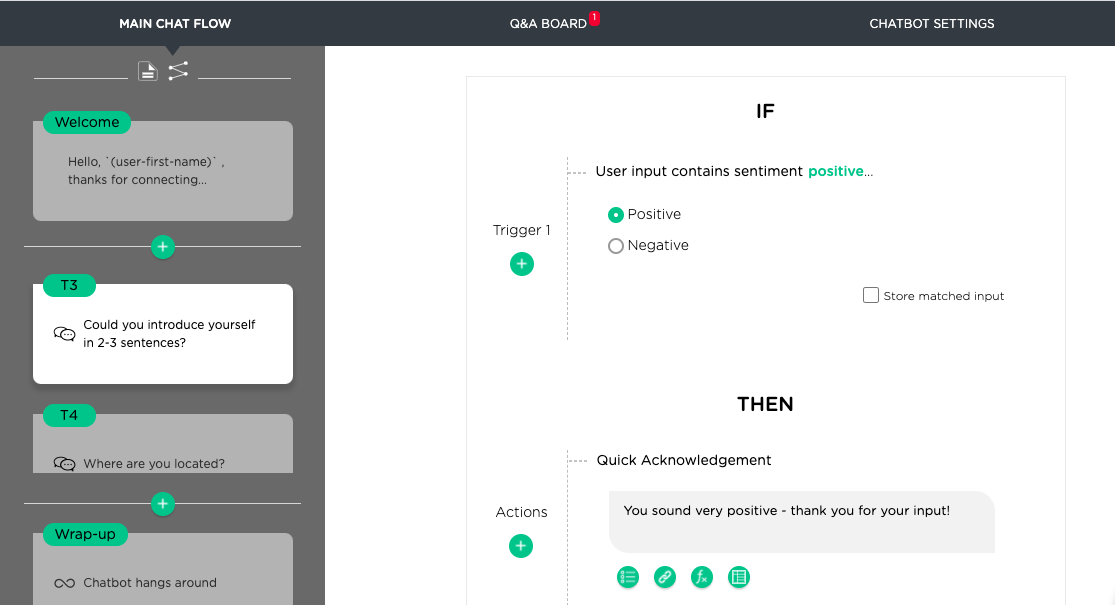}
         \caption*{(b)}
         \Description{A screen shot of Juji dashboard}
     \end{minipage}
     \hfill
     \begin{minipage}[b]{0.32\textwidth}
         \centering
         \includegraphics[width=1.07\textwidth]{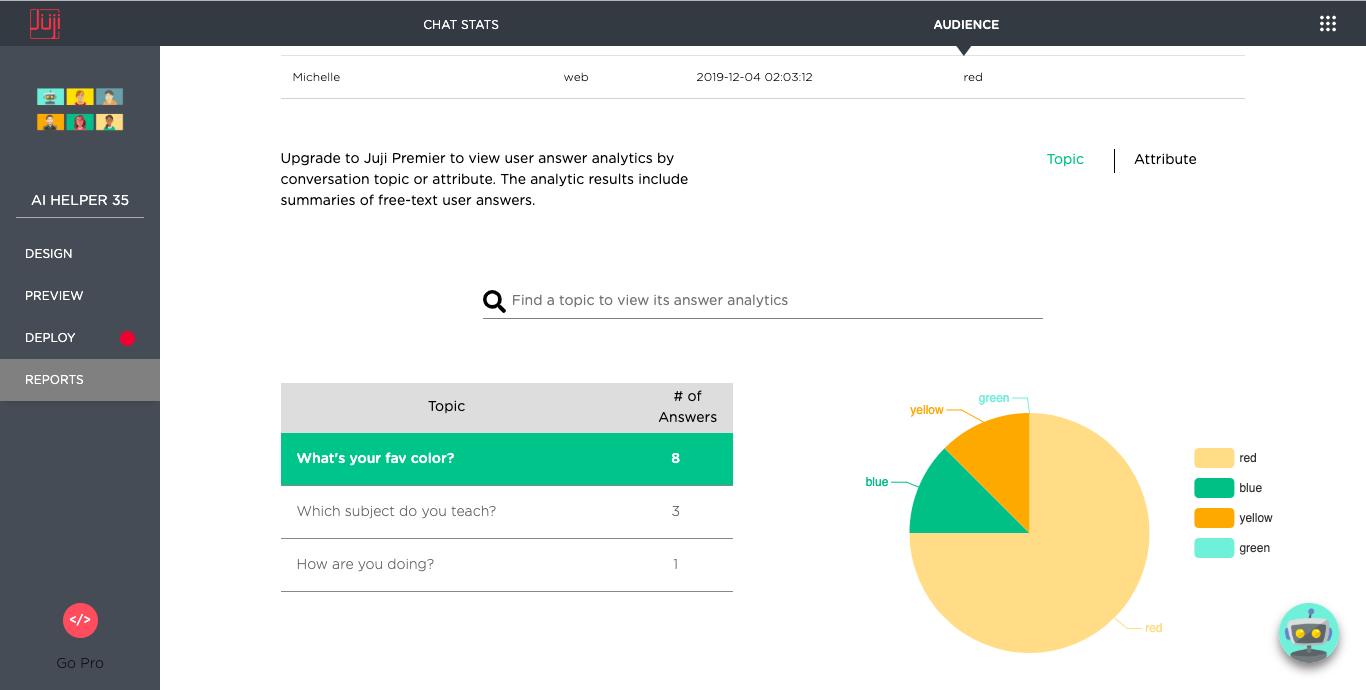}
         \caption*{(c)}
         \Description{A screen shot of Juji dashboard for designers to access summarized audience responses with visualizations such as pie charts.}
     \end{minipage}
     \hfill
     \caption{Example chatbot customizations supported by Juji. (a)
       Customizing a chat flow and a chatbot question. (b) Customizing
       a chatbot response based on user sentiment. (c) The report dashboard that displays interviewee responses visually by Juji.}
      \label{fig:jujiinterface}
\end{figure*}

\section{Study Platform - Juji}
As mentioned in Section 2.2, there are three types of chatbot
platforms. Although we could build \textit{iChatProfile} on top of any
chatbot platform, we decided to build it on Juji for three main
reasons. 

\subsection{Supporting Interview Chatbots}

First, recent studies show that other researchers have used Juji to build
various interview chatbots, which matches our focus on aiding the
design of effective interview chatbots \cite{Li2018-gd, Xiao2019-ff, Xiao2020-fr, volkel2020trick}. Building and deploying an interview chatbot on Juji is very similar to
creating a survey on a popular survey platform like SurveyMonkey or
Qualtrics. Designers use Juji's GUI to enter a set of interview
questions and Juji will \textit{automatically} generate a publicly
accessible interview chatbot with a set of default conversational
skills \cite{Xiao2019-hk}. Juji also automatically handles side
talking and keeps a conversation on track to ensure the completion of
an interview \cite{Xiao2020-fr}.

\subsection{Supporting Non-IT Designers}
For non-IT chatbot designers, Juji
relieves them from implementing many needed AI skills of interview
chatbots while providing them with much freedom to customize a
conversational experience. Specifically, Juji offers a graphical user interface (GUI) for chatbot designers to create, deploy, and manage their custom chatbots without coding
\cite{noauthor_Juji}.  Below are some of the common chatbot customizations
supported by Juji GUI.

\smallskip
\noindent \textbf{Customize chatbot questions/messages}.  Not only can designers easily add/edit/delete text-based chatbot questions, but can also customize questions or messages by adding paraphrasing, inserting URLs or images, and using functions in such text (Fig~\ref{fig:jujiinterface}(a)). For example, one can insert a function to retrieve an interviewee's name to personalize a conversation. 

\smallskip
\noindent \textbf{Customize chatbot responses and persona}. Designers can customize chatbot responses to user input by either directly reusing Juji pre-built
conversations \cite{noauthor_Jujidocs} or defining their own if-then
statements. For example, Fig~\ref{fig:jujiinterface}(b) shows such a
customization. It states that if interviewees' responses contain
positive sentiment, the chatbot would then acknowledge such input
accordingly. Additionally, one can customize chatbot persona or the pace of a conversation.

\smallskip
\noindent \textbf{Access interviewee responses}. 
To help designers monitor interview progress and make design adjustments, Juji provides designers with an interactive report dashboard that displays interviewee responses visually {Fig~\ref{fig:jujiinterface}(c)}. These responses are automatically extracted from the interview conversations. Designers can also download all interviewee responses into a CSV file that contains all the question-response pairs collected.  Because non-IT designers can use Juji to create working interview chatbots, Juji platform is suitable for testing whether 
our approach can assist any chatbot designers to improve their chatbot iteratively.

\subsection{Supporting Easy Integration and Public Access}

\noindent Lastly, the extensibility of
Juji makes it easy for us to build and integrate
\textit{iChatProfile}. Specifically, Juji provides APIs for developers
to access chatbot services and extend Juji chatbots with third-party
functions \cite{noauthor_Jujiapi}. Moreover, Juji is publicly
available and provides an easy access for our study participants and
also for others who wish to replicate or extend our work.
\section{Formative Study and Derived \textit{iChatProfile} Design Guidance}

To guide the development of \textit{iChatProfile}, we conducted a study to first identify the types of design assistance desired by chatbot designers. 

From a public
university, we recruited five students (3 males and 2 females, age ranges from 18 to 31) who
were interested in building chatbots. None of them reported any prior
chatbot design experience. Three of them interacted with
conversational agents like Siri or Amazon Alexa. Our study was a semi-structured, face-to-face online interview. Each
interview lasted about an hour and each participant received \$20 for
their time. At the beginning of each interview, the participants were
asked about their past chatbot design experience.  The participants
were then given a 15-minute tutorial of Juji. They were also
encouraged to try different Juji features and get themselves familiar
with the Juji GUI. After the tutorial, the participants were asked to
use Juji to design an interview chatbot that elicits user input about
the COVID-19 pandemic. They were given a list of questions on this
topic (Table~\ref{tb:questiontable}). 

\begin{table}[htb]
\renewcommand\arraystretch{1.2}
\caption{Interview Questions Used to Build a Chatbot}
\begin{tabular}{ll} 
      \hline
      Q1 & How are you feeling today?      \\
      Q2 & Where you are located?   \\
      Q3 &  What do you do outside work?       \\
      Q4 & What are the challenges you currently face?\\
      Q5 & What do you think you can do to help w/ this pandemic? \\
      \hline
  \end{tabular}
~\label{tb:questiontable}
\end{table}

We selected this set of interview questions for three reasons. First, we wanted to ensure the practical value of our tool development effort, which is to help designers build interview chatbots for real-world uses (e.g., practical user research). Second, we wanted the interview questions to appeal to a wide audience who would be interacting and evaluating the designed chatbots. Third, COVID-19 is a pressing topic that satisfies both criteria.

In this study, we intentionally did not set any specific design
requirements because we wished to observe what the participants would
do and the challenges they would face. Each participant was allotted
30 minutes to design their interview chatbot. The allotted time was determined based on the results of a pilot study where 3 participants could accomplish such a design task well within 30 minutes. After completing the task,
participants were interviewed to discuss the types of design help they had hoped to receive during their design process. We transcribed the audio conversations from these discussions. 

We followed qualitative analysis methods and the grounded theory
\cite{muller2014curiosity, Xu2014-fc} to code the participant interview data. During the first pass, two coders individually reviewed
and coded participants' responses. They then met and discussed their
respective codes to identify common themes and reconcile
differences. Below we report the main findings, which influenced the
design of our computational framework for evaluating the performance
of interview chatbots as well as the development of
\textit{iChatProfile}.

\subsection{Two Types of Design Assistance Wanted}
During the participant interviews, all participants expressed the importance of receiving design assistance. Our analysis also revealed two main types of design assistance that the participants wanted.
The first type (T1) is objective, \textit{quantitative} feedback on their existing chatbot
design that could help designers understand the chatbot deficiencies and point them to the right directions to improve their chatbot.  Almost all participants expressed the need for
receiving such feedback on their chatbot. For example, one
participant mentioned that \textit{"I hope to receive some feedback
  telling me the exact score my chatbot will get ... Just like those
  user ratings on the website of Alexa skills."}.

The second type (T2) is design suggestions for improving a chatbot. Almost all participants
expressed that they still would not know what to do even if a
quantitative evaluation was available.  For example, one participant
stated \textit{"I am really new to this (interview chatbot design). I
  am afraid even I was told this part should be improved, I still
  don't know how. More specific design suggestions would be of great
  help."}  This indicates that designers also wish to receive concrete and actionable
design suggestions that could guide them to improve a chatbot. 

In addition to obtaining design guidance, the participants also expressed the need of viewing relevant conversation examples in the chatbot "debugging" process. For example, one participant stated \textit{"When doing chatbot response customization, I thought a lot about the wording choice since we all know that everyone’s having a hard time during this pandemic. I hoped my chatbot can always be empathetic but I have to admit it might not be the case due to so many different real-world cases."} In such a case, providing designers with the actual conversation fragments (evidence) might help them better grasp the conversation situations and improve their chatbot. In other words, augmenting design suggestions with real conversational examples would also be helpful.

\subsection{\textit{iChatProfile} Design Guidance}
Based on the desired design assistance, we derived three design goals of \textit{iChatProfile} so it can fulfill designers' needs: 
\begin{itemize}
\item {Evaluate the performance of an interview chatbot \textit{quantitatively}
  and present the evaluation results in a structured way. (G1)}
\item {Provide specific, actionable design suggestions based on the
  evaluation results to help a chatbot designer improve the chatbot. (G2)}
\item {Augment design suggestions with evidential conversation examples
  to guide a chatbot designer to make design choices. (G3)}
\end{itemize}

In addition to the three goals directly determined from the findings of our formative study (G1 from T1, G2+G3 from T2), we derived another two criteria to guide the implementation of \textit{iChatProfile} for practical purposes:
\begin{itemize}
\item{\textbf{Adoption}. Ensure that non-IT experts can easily utilize \textit{iChatProfile}. (C1) }
\item{\textbf{Compatibility}. Ensure that \textit{iChatProfile} can be utilized regardless which chatbot platforms are used for designing chatbots. (C2)}
\end {itemize}
We derived C1 as \textit{iChatProfile} is intended to help chatbot designers especially those with no AI/NLP expertise to design, evaluate, and improve interview chatbots. As a result, our effort will help democratize the applications and adoption of
conversational AI. The purpose of C2 is to enable \textit{iChatProfile} to be used with a wide range of chatbot platforms beyond Juji and benefit more designers. Following the goals (G1-G3) and the criteria (C1-C2), we designed \textit{iChatProfile} as discussed in section 6.

\section{Computational Framework for Quantifying Interview Chatbot Effectiveness}

Since our formative study indicated that chatbot designers wish to obtain certain quantitative feedback on the performance of their existing chatbot (T1), we first formulated a computational
framework that quantitatively measures the effectiveness of such an interview
chatbot from multiple aspects. The framework aims at achieving two
goals: 1) providing quantified insights into the performance of an
interview chatbot; 2) using such insights to provide specific and
practical design suggestions for improving the chatbot.

Based on the previous work on assessing human interviews
\cite{Presser2004-pa, chen2011finding, garbarski2016interviewing,
  halbesleben2013evaluating, hess2013linking, allmark2009ethical},
communication theories for conducting effective interviews
\cite{paul1975logic, Xiao2019-hk}, and evaluating chatbot effectiveness
\cite{see2019makes, Guo2018-gm, Xiao2020-fr, Deriu2019-wo,
  Zhou2018-au}, we formulated a set of performance metrics to quantitatively
assess the effectiveness of an interview chatbot from three main
dimensions: elicitation ability, user experience, and ethics.

To ensure both the coverage and practicality of chatbot evaluation, we
used four criteria to choose our metrics. First, we selected only
metrics that can be used to generate design suggestions and help
designers improve an interview chatbot. Second, we chose metrics to
measure both a chatbot's abilities to complete an interview task
effectively (\textit{elicitation abilities}) and a user's experience
with the chatbot (\textit{user experience}) because an ideal interview
chatbot should be able to complete interview tasks while delivering
a satisfactory user experience. Moreover, we included metrics to
evaluate the \textit{ethics} of an interview chatbot because such a
chatbot might engage with a user in a conversation on private and
sensitive topics \cite{Xiao2019-hk, allmark2009ethical}. Third, we
chose metrics to measure the performance of an interview chatbot both
``locally'' (interview question level) and ``globally'' (interview
level). For example, the metric \textit{informativeness} measures the
amount of information conveyed by user responses to each interview
question, while the metric \textit{user sentiment} measures a user's
overall interview experience with a chatbot. This is to ensure specific design
suggestions can be generated to help designers improve a chatbot
question by question (locally), while providing designers with an assessment of the overall interview experience (globally). Fourth, we chose only
metrics that can be easily obtained/computed from available data
(e.g., chat transcripts). This is to facilitate real-time, automatic assessment of chatbot performance and design suggestion generation. Table ~\ref{tb:evaluationframework} summarizes all the metrics.

\begin{table*}[htb]
\renewcommand\arraystretch{1.1}
\caption{Metrics for evaluating the performance of interview chatbots.}
\centering
\resizebox{1.9\columnwidth}{!}{\begin{tabular}{l!{\color{black}\vrule}l|llll}
\hline 
\multicolumn{2}{c}{Dimension}                   & \multicolumn{1}{c}{Metric}                                            & \multicolumn{1}{c}{Synopsis}                                          & \multicolumn{1}{c}{Category} &                                        \\  
\hline
\multirow{4}{*}{\begin{tabular}[c]{@{}l@{}}Elicitation~\\Ability\end{tabular}} &  \multicolumn{1}{l!{\color{black}\vrule}}{\begin{tabular}[c]{@{}l@{}}Response~\\Quality\end{tabular}} & Informativeness & How much information a user reponse conveys  & question level  & \cite{grice1975logic, Xiao2019-hk}                   \\ 
\cline{2-6}
& \multirow{3}{*}{\begin{tabular}[c]{@{}l@{}}User \\Engagement\end{tabular}}&
Response Length                            & The word count in a user's text input & question level  & \cite{Xiao2019-hk}           \\
&                                                            & Engagement Duration & How long a user engages with the chatbot                                  & question level  & \cite{Guo2018-gm, Xiao2019-hk}                  \\
&    & Completion Rate  & \begin{tabular}[c]{@{}>{}l@{}}The percentage of users complete\\an interview question or interview\end{tabular}                                               & \begin{tabular}[c]{@{}>{}l@{}}question \& \\ interview level \end{tabular}  & \cite{conrad2005interactive, chen2011finding, halbesleben2013evaluating}\\ 
\hline
\multicolumn{2}{c|}{\multirow{5}{*}{User Experience}}                           & User Satisfaction Rating & \begin{tabular}[c]{@{}l@{}}A user's satisfaction with the chat\end{tabular}   & interview~ level  & \cite{rodden2010measuring} \\
\multicolumn{2}{c|}{}                                                          & User Trust Rating                        & A user's trust in the chatbot        & interview~ level  & \cite{Li2018-gd, Zhou2018-au, zhou2019trusting} \\
\multicolumn{2}{c|}{}                                                          & User Sentiment    & \begin{tabular}[c]{@{}l@{}}A user's sentiment towards the chat \\ experience\end{tabular}
& interview~ level  & \cite{chen2011finding} \\
\multicolumn{2}{c|}{}                                                         & Level of Empathy                         & The level of empathy expressed by the chatbot~  & question level       & \cite{Zhou2018-au, rashkin2018towards}    \\
\multicolumn{2}{c|}{}                                                          & Repetition Rate   & {\begin{tabular}[c]{@{}l@{}}How much the chatbot repeats itself\end{tabular}}
& question level     & \cite{see2019makes} \\ 
\hline
\multicolumn{2}{c|}{\multirow{2}{*}{Ethics}}                                    & Hate Speech Rate                         & \begin{tabular}[c]{@{}>{}l@{}}How much hate\\speech is contained in chatbot utterances\end{tabular}            & question level                            & \cite{henderson2018ethical, alexaguide}  \\
\multicolumn{2}{c|}{}                                                           & Privacy Intrusion Rate   & \begin{tabular}[c]{@{}l@{}}How much private or sensitive information\\is elicited\end{tabular} & question level                     & \cite{henderson2018ethical, alexaguide}  \\
\hline
\end{tabular}}
  
~\label{tb:evaluationframework}
\end{table*}

\subsection{Elicitation Ability}

The primary task of interview chatbots is to elicit high-quality
responses from participants. Existing literature shows that the
success of an interview is often determined by two aspects: the
elicited response quality and level of user engagement
\cite{garbarski2016interviewing, chen2011finding, o2008user, see2019makes, Guo2018-gm, Xiao2020-fr}. 
We thus model an interview
chatbot's elicitation abilities from two sub-dimensions: response
quality and user engagement. While \textit{response quality} directly
assesses the quality of user responses to an interview question, the
level of \textit{user engagement} quantifies how much a participant is 
engaged with a chatbot from multiple aspects (e.g., how long an
engagement is).

\subsubsection{Response Quality}
We developed a metric to evaluate the quality of user interview responses.

\vspace{\baselineskip} 
\noindent \textbf{Informativeness}. This metric indicates how much
information a user's text response contains. Similar to the metric
used in \cite{Xiao2019-hk}, we measure a word's surprisal---a
word's rareness appearing in modern English
\cite{Wikipedia_contributors2020-oy}. To enable easy reuse of our
metric regardless which English dictionary is used, we compute the informativeness of a user input (\textit{U})
as a sum of the normalized surprisal of each word in \textit{U}:
\begin{equation}
Informativeness(U) = \sum_{n=1}^{N} \textstyle\frac{surprisal(word_n) - min\_surprisal}{max\_surprisal - min\_surprisal} 
\end{equation}
Here \textit{min\_surprisal} and \textit{max\_surprisal} are the minimum and maximum of surprisal, computed among all words in the vocabulary. \textit{N} represents the word count within the response. Currently, we use the Wikipedia Corpus \cite{noauthor_undated-pv} to estimate word frequency.

This metric (e.g., a low \textit{informativeness} score) can signal
designers that there are potential issues with an interview question. For example, a question might be too broad and follow-up questions are needed to elicit more informative responses.

\subsubsection{Level of User Engagement}

In the context of interviews, the \textit{level of user engagement}
measures a user's behavior during an interview
\cite{garbarski2016interviewing}. Specifically, we have defined a set
of metrics to assess a respondent's behavior when engaging with an
interview chatbot. 

\noindent \textbf{Response Length}. This metric computes the
  word count of a respondent’s free-text response to an interview
  question. We chose this metric because previous work indicates that
engaged respondents are more willing to give long responses
\cite{Xiao2019-hk}. Designers can use this metric to
  gauge their chatbot performance and to make
corresponding design improvements (e.g., adding follow-up
  questions or changing a yes/no question to an open-ended question to
  elicit longer responses).

\noindent \textbf{Engagement Duration}. This metric indicates how long a
participant is willing to engage with an interview question. Although
engagement duration alone does not signal the quality of user
responses \cite{Xiao2019-hk}, we hope to use it as an indicator of potential issues
with an interview question. For example, if the engagement duration of 
a particular open-ended interview question is exceedingly short, it might signal
that the question is too narrow and needs to be rephrased to encourage more open and 
longer engagement.

\noindent \textbf{Completion Rate}. This metric computes the percentage of participants
completing an interview question or an entire interview. It is a commonly used metric to
measure the effectiveness of an interviewer \cite{chen2011finding,
  halbesleben2013evaluating}. To better help designers improve their
chatbots question by question (see Section 6), we compute the
completion rate for each interview question (\textit{Q}) by counting the number of users who completed the question ($C_q$) and the number of users who responded to the question ($T_q$):
\begin{equation}
CompletionRate(Q) = {C_q}/{T_q}
\end{equation}
\noindent For the first interview question (when n = 1), we directly use the        
number of participants as the numerator.

\noindent This metric (e.g., a low completion rate) can be used to signal
potential issues related to an interview question (e.g., too vague) or the chatbot's inability to handle user responses to the question (e.g.,
user's expressed unwillingness to answer this question). A low interview-level completion rate could also reflect potential issues with an interview (e.g., too many questions). Corresponding
design suggestions can then be made to help the designers improve the
chatbot.

\subsection{User Experience}
Informed by literature in interaction design \cite{rodden2010measuring} and 
interview design \cite{chen2011finding, halbesleben2013evaluating},
we proposed five metrics to measure a user's experience with an interview chatbot. 
\smallskip

\noindent \textbf{User Satisfaction Rating}. This metric is directly computed
from participants' ratings of their chatbot interview experience. This
rating can be easily obtained: when piloting an interview chatbot,
a question like \textit{"How satisfied are you with the interview experience?"}
can be added at the end of an interview session for a participant
to report their level of satisfaction.

\noindent \textbf{User Trust Rating}. This metric measures participants'
perceived trust in an interview chatbot. Trust is important because it 
affects participants' willingness to share information
\cite{Li2018-gd}. Similar to obtaining the user satisfaction rating, a
question like 
\textit{"How much do you trust this chatbot? Please rate it on a scale of 1 to 5"} can be 
added at the end of an interview when piloting an interview chatbot.

\noindent \textbf{User Sentiment}. This metric evaluates participants' sentiment
toward an interview chatbot since such a metric is widely used to
measure user satisfaction with interviews/surveys
\cite{chen2011finding}. To obtain user sentiment, one can elicit
participants' rationale (why) when eliciting their satisfaction rating
and trust rating during pilot interviews. Currently, we use the Vader model \cite{gilbert2014vader} to perform sentimental analysis on the
collected users responses, and compute the percentages of positive,
neutral and negative responses. 

\noindent \textbf{Level of Empathy}. This measures the level of
empathy an interview chatbot has since research shows that an
empathetic chatbot is able to elicit higher quality responses
\cite{Xiao2020-fr}. Currently, we compute the level of empathy by the
frequency of empathetic words used by a chatbot. Specifically, given a
conversation segment associated with interview question \textit{Q}, we
normalize the number of empathetic words ($E_c$) over the total number of words within chatbot utterances ($T_c$) in this segment: 
\begin{equation}
LevelOfEmpathy(Q) ={E_c}/{T_c}
\end{equation}

\noindent We extracted the empathetic words from \textit{EmpatheticDialogues}
by identifying top 15 content words from
each of its 32 emotion categories \cite{rashkin2018towards}.  This metric can help designers identify chatbot
responses that lack of empathy and make corresponding improvements.
Although more sophisticated algorithms can be used to measure empathy \cite{Zhou2018-au}, we opted for the current approach that requires little
training so that others can easily adopt it even without AI/NLP
expertise or training data.

\noindent \textbf{Repetition Rate}. This metric computes
  the frequency an interview chatbot has to repeat itself during an
  interview.

\noindent We include this metric for two reasons. First, repetition affects the
quality of a dialogue system, which in turn influences user experience
\cite{see2019makes}. Second, repetition may signal a chatbot's
inability to handle certain user input. For example, a chatbot might
not be able to handle unexpected user input and have to re-ask
an interview question \cite{Xiao2020-fr}. Currently, given a
conversation segment associated with an interview question
(\textit{Q}), we normalize the number of the repeated bi-grams ($R_c$)
over the total number of bi-grams ($T_c$) of chatbot's utterances:
\begin{equation}
RepetitionRate(Q) = {R_c}/{T_c}
\end{equation}

\subsection{Ethics}

An interview chatbot may engage participants in a conversation on
private or sensitive topics or the participants may voluntarily offer
private and sensitive information \cite{Li2018-gd, Zhou2018-au, alexaguide, henderson2018ethical}. It thus is important to build ethical chatbots that
respect participants as well as protect their privacy. We thus have
developed two metrics to evaluate the ethics of an interview chatbot.

\smallskip
\noindent \textbf{Hate Speech Rate}. This metric assesses how much an
interview chatbot includes hate speech in its utterances. Such
assessment becomes even more important if a chatbot uses
auto-synthesized responses as what Tay was using
\cite{noauthor_Tay}. Currently, we use an automated hate speech
detection algorithm to compute the hate
speech rate \cite{davidson2017automated}. This metric can help chatbot designers be better aware of
a chatbot's built-in AI capabilities and correct a chatbot's behavior
if needed.

\noindent \textbf{Privacy Intrusion Rate}. This metric evaluates how
much an interview chatbot elicits private or sensitive information
from a participant (e.g., password or social security
number). Currently, for each interview question, we first identify
``sensitive'' words/phrases appearing in chatbot utterances or user responses using Google's
Data Loss Prevention (DLP) API \cite{noauthor_DLP}. These words, such as a social
security number, might risk a user's privacy. Given a conversation
segment associated with an interview question (\textit{Q}), we compute
the rate as follows:
\begin{equation}
PrivacyIntrusion(Q) = {S_c}/{T_c}
\end{equation}
\noindent Here $S_c$ is the count of sensitive words appearing in the user responses and $T_c$ is the total word count in user responses.

\noindent To better protect a user's privacy, chatbot designers can use this
metric to curb an interview chatbot from eliciting such information or
reminding a user of not giving up such information
unnecessarily during an interview.

\section{iChatProfile}

To help chatbot designers evaluate and improve an interview chatbot
iteratively, we have developed a tool called \textit{iChatProfile}, following the design goals and design criteria summarized in Sec 4.2. It
automatically computes the metrics (Table~\ref{tb:evaluationframework}) to assess the performance of an interview chatbot and generates a chatbot profile. Based on the
profile, \textit{iChatProfile} also automatically generates a set of design suggestions for
improving the chatbot. 

\begin{figure}
     \centering
     \begin{minipage}{0.45\textwidth}
         \centering
         \includegraphics[width=\textwidth]{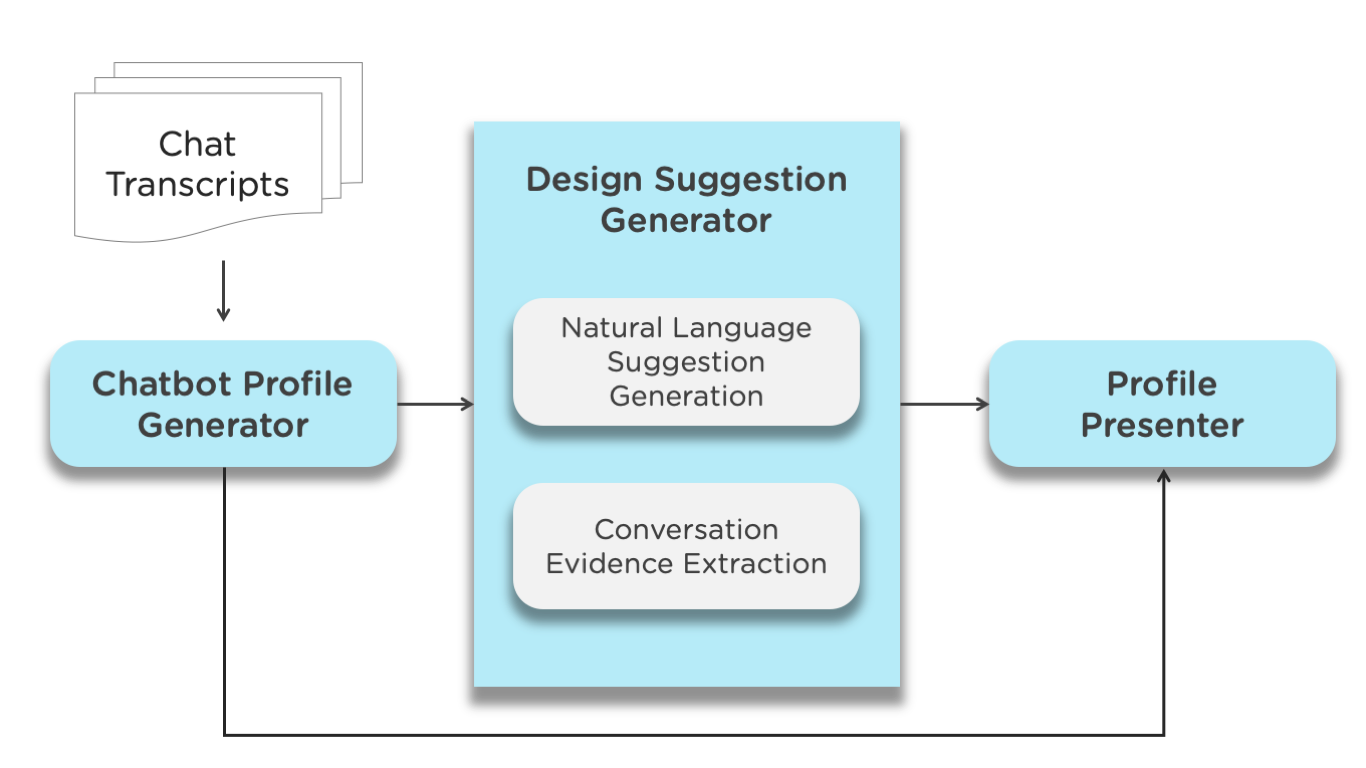}
         \caption*{(a)}
         \Description{The diagram is a flow chart with four main components. Each component is connected with an arrow to indicate the flow. }
     \end{minipage}
     \hspace{0.025\textwidth}
     \begin{minipage}{0.45\textwidth}
         \centering
         \includegraphics[width=\textwidth]{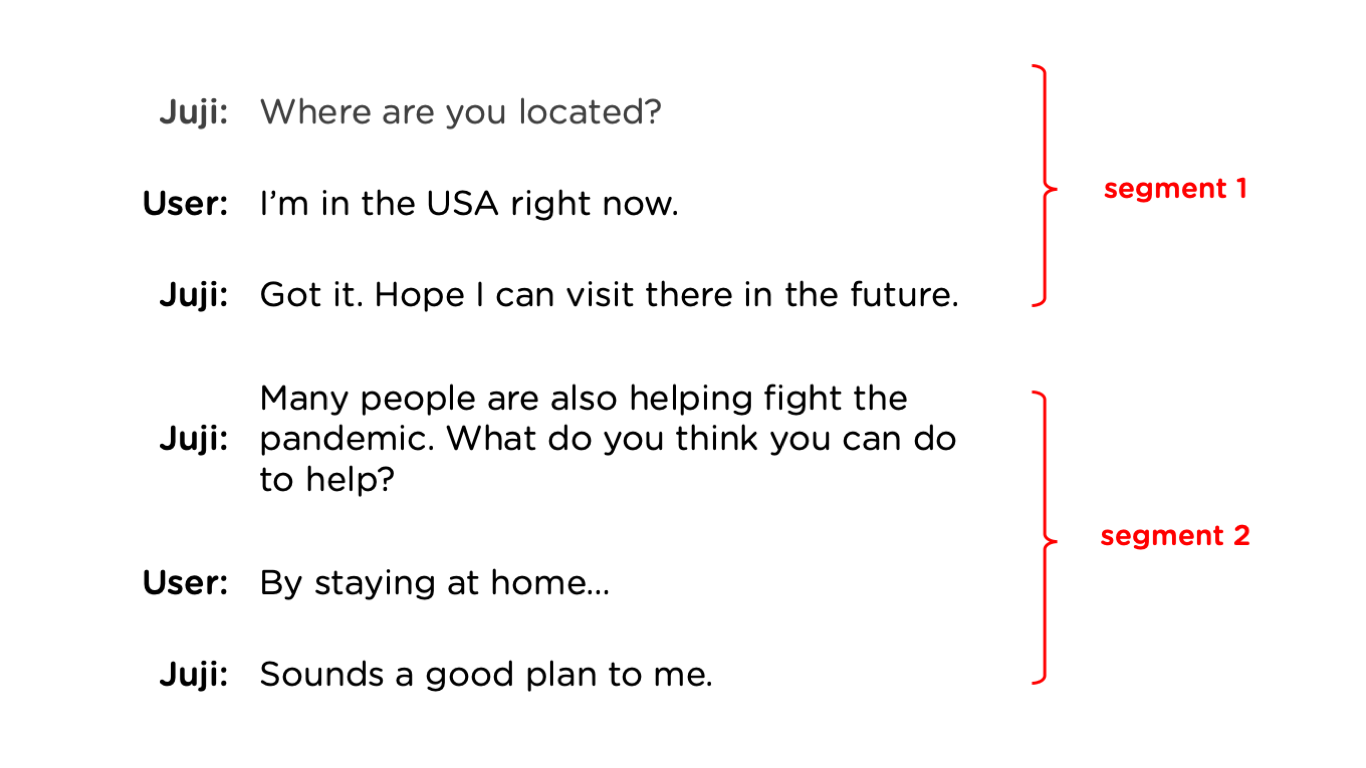}
         \caption*{(b)}
         \Description{The chat transcript contains two segments}
     \end{minipage}
     \caption{An overview of \textit{iChatProfile}. (a) key components. (b) an example chat transcript and its conversation segments.}
     \label{fig:ichatprofileoverview}
\end{figure}

\subsection{System Overview}

As shown in Fig~\ref{fig:ichatprofileoverview}(a), \textit{iChatProfile} consists of
three key components: chatbot profile generator, design
suggestion generator, and profile presenter. The
\textit{chatbot profile generator} takes a set of chat transcripts as
input and automatically computes all the chatbot performance metrics
(Table ~\ref{tb:evaluationframework}) to generate a chatbot profile. In general, chat
transcripts are the results of the live chats between an interview
chatbot and its pilot/testing users. Once a chatbot profile is
created, the \textit{design suggestion generator} automatically
generates a set of specific suggestions with conversation evidence
  for improving the chatbot. The chatbot profile, the design
  suggestions, and the conversation evidence are then assembled
  together by the \textit{profile presenter} and displayed in a visual dashboard for easy comprehension
  (C1). To make \textit{iChatProfile} easily work with any chatbot
  platforms, it is implemented as an independent tool and takes only chat transcripts as its input (C2).

\subsection{Profiling Interview Chatbots}

Given a set of chat transcripts, \textit{iChatProfile} automatically
computes all the metrics mentioned in Section 5 to assess the
performance of an interview chatbot. Each chat transcript is first
segmented by interview question and each segment consists of one or
more conversation turns (Fig~\ref{fig:ichatprofileoverview}(b)). Each
metric (e.g., \textit{response length}), except \textit{completion
  rate} and \textit{user sentiment}, is first computed/extracted per
transcript (user) and all the scores are then averaged across all
transcripts (users). The \textit{completion rate} and \textit{user
  sentiment} are directly calculated from all the transcripts (e.g.,
Formula 2).

\subsection{Generating Design Suggestions}

Given the computed performance metrics, \textit{iChatProfile}
automatically generates a set of design suggestions using a rule-based
approach. To make each design suggestion actionable, we formulate
rules based only on the question-level evaluation metrics (e.g.,
\textit{informativeness}). Such a design suggestion can be used by a
chatbot designer to further customize and tweak chatbot behavior
around a specific interview question. In contrast, it is difficult for
designers to act upon an interview-level metric, such as \textit{user
  satisfaction rating}, although it informs the designers the overall
performance of a chatbot.

Below is an example rule. It states that if the computed
\textit{repetition rate} for an interview question (\textit{Q}) is
above a certain threshold, it then uses a template to generate a set
of design suggestions that could reduce repetitions and improve user
experience.

\begin{algorithm}
\begin{algorithmic}[1]
\IF{\textit{repetition rate(Q)} > \textit{threshold}}
\STATE generate-design-suggestions (reduce-repetition-template)
\ENDIF
\end{algorithmic}
\end{algorithm}

In our current implementation, the default thresholds are determined
by the corresponding metric scores of the opening question (Q1). This is
because a recent study shows that the conversation around the very
first question could be used as a good indicator
\cite{Xiao2019-hk}. The only exception is for \textit{hate speech
  rate}, where the threshold is set to 0. It means that if any hate
speech is detected, design suggestions will be
generated. Additionally, the thresholds can also be defined by
designers themselves based on their needs. 

Once a rule is triggered, \textit{iChatProfile} automatically
generates actionable design suggestions in two steps. First, it uses a
template-based approach to generate design suggestions in natural
language \cite{Najafian2020-az}. Second, it automatically extracts
relevant conversation fragments as evidence to substantiate the
generated design suggestions.

\begin{table*}[hbt!] 
\renewcommand\arraystretch{1.3}
\centering
\caption{Metric-based chatbot design guidelines.}
\resizebox{2\columnwidth}{!}{\begin{tabular}{lll} 
\hline
 \multicolumn{1}{c}{Design guideline}      &                                                                                          \multicolumn{1}{c}{Metric}                                                                                                                                                                                                                                                                                                                  &                   \\ 
\hline
 Add polite probings
  and explanations to the question & Informativeness, Completion Rate  & \cite{de2016handling, oudejans2010using}                                    \\
  Add customizations to show the chatbot is actively listening & Informativeness, Engagement Duration & \cite{Xiao2020-fr} \\
  Set a minimum response length to handle short user input & Response Length, Engagement Duration & \cite{noauthor_Jujidocs} \\ 
   Add customized chatbot responses to handle user digressions & Response Length, Engagement Duration, Repetition Rate & \cite{Xiao2019-hk, Xiao2020-fr, see2019makes} \\
  Reword the question to make it more acceptable to users & Completion Rate, Repetition Rate & \cite{alexaguide, de2016handling, heerwegh2007personalizing}\\
   Personalize the chat experience, e.g., addressing users their names & Completion Rate &\cite{alexaguide, heerwegh2007personalizing, noauthor_Jujidocs}\\
  Add default empathetic chatbot responses to handle unknown user input & Level of Empathy & \cite{alexaguide, Zhou2018-au, noauthor_Jujidocs} \\ 
   Customize chatbot responses to give empathetic feedback on user input & Level of Empathy & \cite{alexaguide, Zhou2018-au} \\
  Remove all the hate or offensive speech & Hate Speech Rate & \cite{alexaguide}\\
   Avoid asking private or sensitive information without user consent & Privacy Intrusion Rate & \cite{alexaguide} \\ \hline

\end{tabular}}
~\label{tb:designguideline}
\end{table*}
\subsubsection{Template-based Natural Language Generation}

For each metric, we have defined a template that contains one or more
design guidelines for improving a chatbot
(Table~\ref{tb:designguideline}). These design guidelines are
formulated based on previous research findings and commercial product
design guidelines (Alexa, Google Home and Juji) for improving
interview quality and user experience \cite{de2016handling,
  oudejans2010using, conrad2005interactive, see2019makes, Zhou2018-au,
  Xiao2020-fr, alexaguide, GuideG, heerwegh2007personalizing,
  noauthor_Jujidocs}.  For example, there are two
  guidelines on improving the metric \textit{informativeness}: one is
  to better articulate or explain an interview question to minimize
  ambiguity, while the other is to improve a chatbot with active
  listening skills to make users feel heard \cite{Xiao2020-fr}.

Given a template, it takes two steps to generate design suggestions in
natural language: document planning and surface realization
\cite{Najafian2020-az}. In document planning, we define the content to
be conveyed in four parts: (a) the design guideline (\textit{D}), (b)
the corresponding interview question (\textit{Q}), (c) the
corresponding metric (\textit{M}), and (d) an explanation on why the
design guidelines are given. In surface realization, we generate
natural language statements by a template: \textit{"For question Q, do
  D because metric M is Z"}. Here \textit{Z} is either "too low" or
"too high", depending on which metric value triggers the
generation. We have used a python library
SimpleNLG\cite{noauthor_simplenlg} to automatically generate
grammatically correct natural language sentences. The library helps
organize basic syntactic structure (e.g., tense) and sentence elements
(e.g., punctuation).

Using the example rule mentioned above, assume that the computed
\textit{repetition rate} for the interview question \textit{"Where are
  you located?"} exceeds the threshold, signaling potential issues
around this interview question. This triggers \textit{iChatProfile} to
generate a set of design suggestions as shown in Fig~\ref{fig:3}(b).

\subsubsection{Conversation Evidence Extraction}

To act on design suggestions, chatbot designers may need more
information to understand the conversation situations. As we
learned from the formative study, it is difficult for designers to
anticipate conversation situations. Continuing the above example,
although a chatbot designer now knows that the interview question
\textit{"Where are you located"} has caused high repetitions, s/he
might not know what caused the repetitions.  In such a case, providing
designers with the actual conversation fragments (\textit{evidence})
might help them better grasp the situations and make chatbot improvements.

Thus, \textit{iChatProfile} automatically extracts relevant
conversation fragments from chat transcripts to give designers more
concrete ideas on how to improve a chatbot. These conversation
fragments are essentially concrete evidence to show chatbot designers
what went wrong. However, such fragments might be too many, which
would not help the designers but overwhelm them. We have thus used
GloVe embeddings \cite{pennington2014glove} and a k-means algorithm
\cite{krishna1999genetic} to select the most representative
conversation fragments in three steps.

Given an interview question (e.g., \textit{``where are you located}'')
and a performance metric (e.g., \textit {repetition rate}),
\textit{iChatProfile} first selects all conversation segments (Fig~\ref{fig:ichatprofileoverview}(b)) that
produced a metric score worse than the threshold. These selected
segments are then encoded by GloVe embbedings. Second, these segments
are grouped into \textit{k} clusters based on their cosine
similarity. Elbow method is used \cite{kodinariya2013review} to find
the optimal number of clusters (\textit{k}). Third,
\textit{iChatProfile} then ranks the clusters by coverage (i.e., the
number of segments in each cluster). Within the top-\textit{K}
clusters, one conversation segment is randomly selected per cluster as
the representative evidence to substantiate the design
suggestions. Currently, \textit{K} is determined by the available
space in the visual dashboard after displaying the design
suggestions. The rest of the conversation segments can also be
accessed through a hyperlink.

Using the above example on the interview question of \textit{``Where
  are you located''}, two clusters are formed, one with the coverage
of 75.0\% and the other 25.0\%. Assuming \textit{K}=2,
\textit{iChatProfile} selects one conversation segment from each
cluster (Table 4 and Table 5).

\smallskip

\noindent\begin{minipage}{.55\linewidth}
   \centering
   \captionsetup{justification=centering}
   \captionof{table}{Conversation Seg-\\ment from Cluster 1}
   \resizebox{0.85\columnwidth}{!}{\begin{tabular}{rlc}\hline
      Juji: & Where are you located?& \\
     User: & You tell me first. & \\
     Juji: & Where are you located? & \\
     User: &  What about you? & \\  \hline
     ~\label{tb:repetitionexample1}
   \end{tabular}}
\end{minipage} \hspace{-0.6cm}
\begin{minipage}{.55\linewidth}
\captionsetup{justification=centering}
    \captionof{table}{Conversation Seg-\\ment from Cluster 2}
   \centering
   \resizebox{0.85\columnwidth}{!}{\begin{tabular}{rlc}\hline
     Juji: &  Where you are located? & \\
     User: &  no  &\\
     Juji: &  Where are you located? & \\
     User: &  I don't want to & \\  \hline
     ~\label{tb:repetitionexample2}
   \end{tabular}}
\end{minipage}

\subsection{Presenting Chatbot Profile and Design Suggestions}

To present a generated chatbot profile, design suggestions, and
relevant conversation evidence, we used Tableau \cite
{noauthor_tableau} to implement a web-based, interactive visual
dashboard  (Fig~\ref{fig:3}). A displayed chatbot profile consists of all computed metrics visualized in various forms depending on the type of information. For example,
\textit{response length} is visualized in a bar chart while \textit{user
  sentiment} is displayed in both a pie chart (showing the percentages
of each type of sentiment) and word clouds (Fig~\ref{fig:3}(a)). The
profile also visually indicates the thresholds that would trigger
design suggestions, which helps designers better understand the
meanings of metric scores and make design decisions. Users can
interact with each metric to view corresponding design suggestions if there is
any. If a performance metric (e.g., \textit{informativeness}) deems to be improved,
\textit{iChatProfile} presents the generated design suggestions and
conversation evidence (Fig~\ref{fig:3}(b)).

\section{Evaluation}

To evaluate the effectiveness of \textit{iChatProfile}, we designed
and conducted a between-subject user study that compared the performance of 10
chatbots designed with or without using \textit{iChatProfile}.

\subsection {Study Method}

Using the same set of interview questions about COVID-19 shown in
Table~\ref{tb:questiontable}, we first built an interview chatbot on
Juji using only Juji's built-in features without making any
customization. After asking all the interview questions, the chatbot
also included questions to elicit user satisfaction rating and trust
rating, as well as their rationale behind each rating. This chatbot
served as our baseline. The baseline chatbot was deployed on the web
to engage with respondents in a live chat. The pilot study collected a
total of 128 chat transcripts. Using these transcripts,
\textit{iChatProfile} automatically generated a chatbot profile and
corresponding design suggestions to improve the baseline chatbot.

Since we wished to compare chatbot performance with and without using
\textit{iChatProfile}, we recruited 10 chatbot designers, who were
randomly divided into two groups, 5 in each group. Each designer
started with a 15-minute tutorial of the Juji platform by watching a
tutorial video and learning several key Juji features (e.g., how to
customize a chatbot's actions). They were given additional time to
play with Juji and get familiar with various design features. Each
designer was then given the baseline chatbot for them to import into their own account so they could preview and improve the baseline. They also had access to the report dashboard and all the interviewee responses extracted from the 128 conducted chatbot interviews as described in section 3.2. They were asked
to describe the good and bad aspects of the baseline chatbot. Next, they were asked to improve the
baseline chatbot along three dimensions: user response quality, user experience, and ethics. They were allowed to use any chatbot customizations (e.g., rewording a question or
customizing a chatbot's reactions to user input) as long as all the
original interview questions and the question order were kept. All the
designers in one group (Group B, w/ \textit{iChatProfile}) were also given \textit{iChatProfile}
to view the generated profile of the baseline chatbot and
corresponding design suggestions, while the other group (Group A, w/o \textit{iChatProfile}) was
not given the tool but only the interviewee responses . We also collected the participants' demographics,
including their gender and age, and their chatbot experience
(chatbot interaction or design experience).

Each designer was allotted about 30 minutes to improve their chatbot. A
post-task interview was also conducted. The designers in  Group w/o  \textit{iChatProfile}  were asked about their design and the challenges/difficulties they faced during their chatbot design process. The designers in Group w/ \textit{iChatProfile}  were asked to describe their design and their experience of using \textit{iChatProfile}. Because of the COVID-19 pandemic, the whole study was conducted online via an 1:1 Zoom meeting. On average, each study session lasted about an hour.

Ten (10) designers from the two groups built a total of 10 chatbots based on the baseline chatbot provided to them.
Each of these chatbots was deployed on the web to engage with respondents in a live chat.
\vspace{-0.2cm}
\subsection {Participants}
All chatbot respondents, including the ones in the pilot study, were
recruited on Amazon Mechanical Turk (MTurk) with an approval rating
equal to or greater than 99\% and located in the U.S. or Canada. 
Each participant was paid \$12.5/hr.

The 10 chatbot designers (6 males, 4 females, ages 20 to 35) were
students recruited from a public university majoring in diverse
disciplines, including Computer Science, Information Science,
Psychology, and Environmental Studies. Two (2) participants reported
prior experience of building chatbots, which five (5) of them reported
prior experience of interacting with conversational agents, like Siri,
Amazon Alexa or Google Assistant. None of them had built interview
chatbots or used Juji. Each participant was paid \$20 for their time.

\subsection{Study Results}

From the 10 deployed interview chatbots, we collected a total of 1349 interview transcripts including the transcripts of incomplete interviews. We kept the incomplete ones because they could indicate the performance of a chatbot. On average, each chatbot interviewed 135 users (135 chat transcripts).  Given their respective interview transcripts, \textit{iChatProfile} computed ten (10) chatbot profiles to characterize the performance of each of the chatbots. We then compared the computed chatbot performance between the 5 chatbots (702 transcripts) designed by the participants in Group w/o  \textit{iChatProfile} and another 5 chatbots by Group w/ \textit{iChatProfile}  (647 transcripts) using the tool.

\subsubsection{\textbf{ANCOVA Analyses}}
Specifically, we performed a series of ANCOVA analyses, which blend
analysis of variance (ANOVA) and regression \cite{keppel1991design},
to examine the effect of with or without using \textit{iChatProfile}
(independent variable) on various chatbot performance metrics
(dependent variables). We ran ANCOVA analyses on every metric in Table~\ref{tb:evaluationframework} except three due to a lack of data: \textit{completion rate} (10 samples), \textit{user sentiment} (10 samples), and \textit{hate speech rate} (no data). Both \textit{completion rate} and \textit{user sentiment} were computed per chatbot (a total of 10 chatbots) and our algorithm did not detect any hate speech in any of the chatbots.

We compared the chatbot performance between Group w/o  \textit{iChatProfile} (702
transcripts) and Group w/ \textit{iChatProfile} (647
transcripts). The assumption check was
conducted to make sure the unequal sample size would not affect the
reliability of the results. In each analysis, the independent variable
was the group and the dependent variable was one of the chatbot
performance metric scores computed. All analyses were controlled for
designers' differences, including their gender and chatbot
experience. We did not control their age because they all are of the
similar age. Whenever applicable (e.g., for \textit{informativeness}
but not \textit{user satisfaction rating}), each analysis was
additionally controlled for respondents' differences---the
corresponding metric score of the first question (Q1). This is because
prior study shows that a respondent's behavior in the opening question
is a significant predictor of his/her behavior in the entire interview
\cite{Xiao2019-hk}.

\begin{table}[htb!]
\renewcommand\arraystretch{1.2}
\centering
\captionsetup{justification=centering}
\caption{Comparison of chatbot performance metrics \\ between Group A (w/o iChatProfile) and Group B (w/ iChatProfile).}
\raggedright
\resizebox{0.77\columnwidth}{!}{
\begin{tabularx}{0.85\linewidth}{l|ll|ll|ll}
    & \multicolumn{2}{l|}{Group B} & \multicolumn{2}{l|}{Group A} &    \multicolumn{2}{l}{Baseline} \\ \cline{2-7}
    
 Metrics & Mean & SD & Mean & SD & Mean & SD \\ \cline{1-7}
Informativeness & 2.160 & 1.362 & 1.905 & 1.645 & 1.843 & 1.061 \\
Response Length & 7.204 & 5.788 & 6.291 & 4.736 & 6.204 & 3.969 \\
Engagement Duration & 2.342  & 0.362 & 1.632 & 0.122  & 0.758 & 0.324 \\ \cline{1-7}
 
Satisfaction Rating & 4.229  & 0.977 & 4.039 & 1.113 & 4.172  & 1.004   \\
Trust Rating & 4.034  & 1.027 & 3.925 & 1.072 & 3.929  & 0.961   \\
 Level of Empathy & 0.018  & 0.039 & 0.015 & 0.043 & 0  & 0   \\ 
 Repetition Rate & 0.016  & 0.003 & 0.018 & 0.006 & 0.018  & 0.047   \\ \cline{1-7} 
Hate Speech Rate & 0  & 0 & 0 & 0 & 0  & 0   \\ 
Privacy Intrusion Rate & 0.217  & 0.182 & 0.204 & 0.194 & 0.208  & 0.181  \\ \cline{1-7}

\end{tabularx}}
  \label{tb:3}
\end{table}

\begin{figure}
\includegraphics[width=0.45\textwidth]{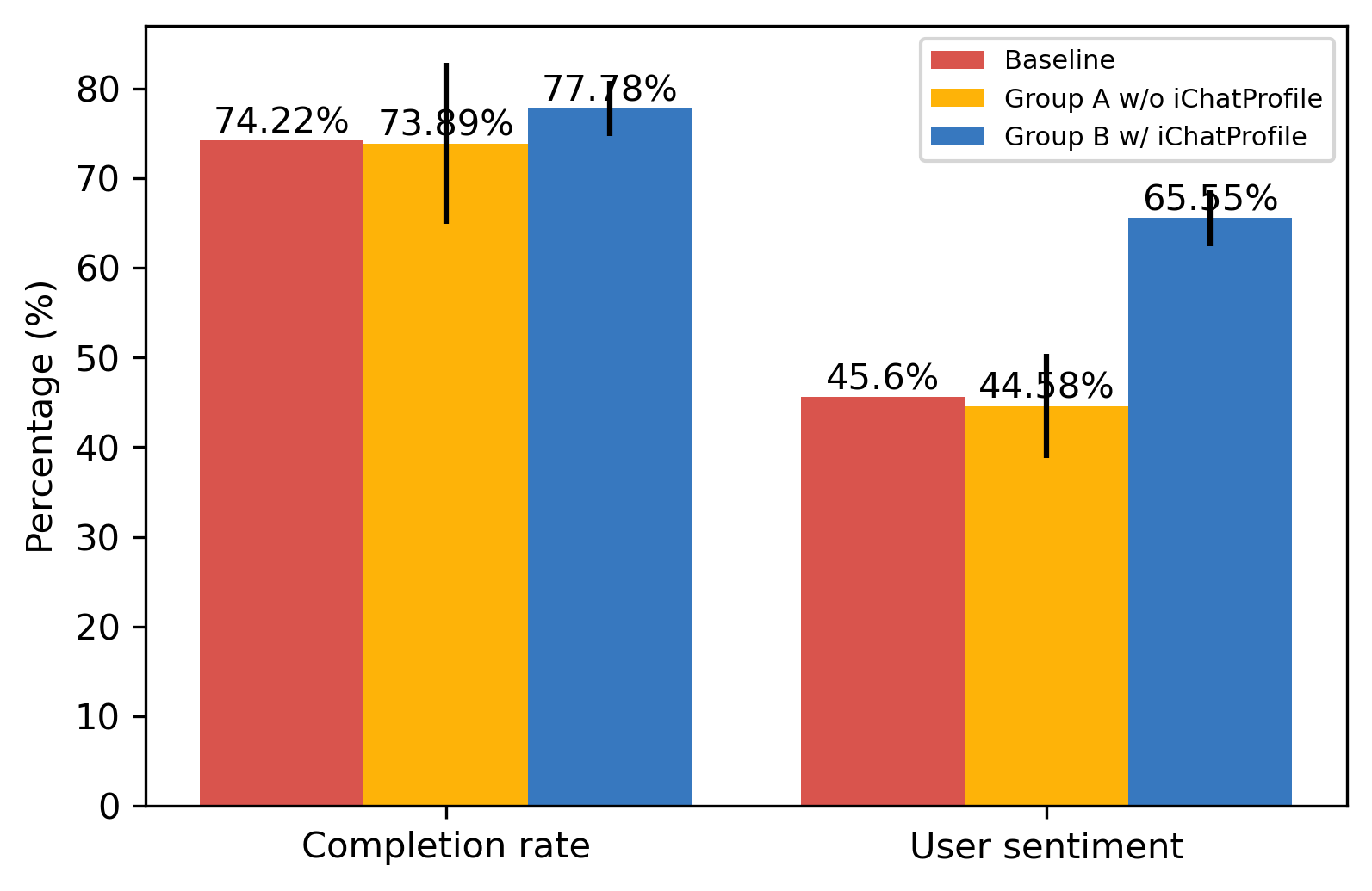}
\caption{Comparison of \textit{completion rate} and
  \textit{user sentiment}.} 
  \Description{This is a bar chart with the performance of group A, group B and baseline side by side.}
\end{figure}

Similarly, we ran ANCOVA analyses to compare each chatbot's performance
metrics between Group w/o  \textit{iChatProfile}  and the baseline (128 transcripts), and between
 Group w/ \textit{iChatProfile} and the baseline, respectively. Bonferroni correction was applied to adjust \textit{p} values. 

\smallskip
Since all the designers were given the same goal to improve a chatbot along three dimensions, our analyses were to answer two questions:

\begin{itemize}
\item{\textbf{RQ1}: Did \textit{iChatProfile} help designers build better interview chatbots?}
\item {\textbf{RQ2}: How did \textit{iChatProfile} help make chatbot design decisions?}
\end{itemize}

\noindent Before running ANCOVA analyses, we also examined the
correlations among all dependent variables. Consistent with prior
findings \cite{Xiao2019-hk}, \textit{informativeness} was not
correlated with \textit{engagement duration}. Moreover,
\textit{engagement duration} did not significantly correlate with any
other metrics except \textit{repetition rate}. We also noted that a
chatbot's \textit {empathy level} significantly correlated with
\textit{informativeness} and \textit{response length}. Intuitively,
this result is sensible since respondents would be more cooperative
with an empathetic chatbot \cite{Xiao2020-fr}.

\begin{table*}[htb!]
\renewcommand\arraystretch{1.2}
\centering
\captionsetup{justification=centering}
\caption{ANCOVA analysis results for chatbot performance metrics between baseline,\\ Group A (w/o \textit{iChatProfile}), and Group B (w/ \textit{iChatProfile}).}

\resizebox{2.0\columnwidth}{!}{
\begin{tabularx}{0.95\linewidth}{l|lll|lll|lll}
    & \multicolumn{3}{c|}{Group B vs. Group A} & \multicolumn{3}{c|}{Group B vs. Baseline} &    \multicolumn{3}{c}{Group A vs. Baseline} \\ \cline{2-10}
    
 Metrics & F & p &$\eta_{p}^2$ & F & p & $\eta_{p}^2$ & F & p & $\eta_{p}^2$\\ \cline{1-10}
Informativeness & 24.28 (1, 1094) & < 0.001 *** & 0.022 & 12.59 (1, 677) & < 0.01 ** & 0.021 & 0.73 (1, 689) & > 0.05 & 0.001 \\
 Response Length & 28.60 (1, 1094) & < 0.001 *** & 0.026 & 7.78 (1, 677) & < 0.05 * & 0.024 & 0.48 (1, 689) & > 0.05 & 0.001 \\
 Engagement Duration & 6.19 (1, 1094)  & < 0.05 * & 0.004 & 9.37 (1, 677)  & < 0.01 ** & 0.007 & 6.47 (1, 689) & < 0.05 * & 0.004 \\ \cline{1-10}

 User Satisfaction Rating & 6.68 (1, 1032)  & < 0.05 * & 0.006 & 5.99 (1, 615) & < 0.05 **  & 0.006 & 4.89 (1, 611) & > 0.05  & 0.007   \\
 User trust rating & 5.95 (1, 1035)  & < 0.05 * & 0.004 & 8.72 (1, 616) & < 0.01 **  & 0.007 & 2.04 (1, 613) & > 0.05 & 0.004\\

 Level of Empathy & 6.07 (1, 1096)  & < 0.05 * & 0.004 & 13.97 (1, 679) & < 0.001 ***  & 0.023 & 13.97 (1, 691) & < 0.001 *** & 0.023 \\
 
 Repetition Rate & 14.30 (1, 1096)  & < 0.001 *** & 0.012 & 8.206 (1, 679) & < 0.05 *  & 0.013 & 0.821 (1, 691) & > 0.05 & 0.013  \\ \cline{1-10}

 
Privacy Intrusion Rate & 0.067 (1,1095) & > 0.05 & 0.000 & 1.378 (1, 678) & > 0.05  & 0.003 & 0.158 (1, 690) & > 0.05 & 0.000 \\ \cline{1-10}

\end{tabularx}}
\vspace{1ex}
     {\raggedright \small \\ \quad \quad \quad \noindent a. \textit{p-value} in this table is  Bonferroni corrected. }
     {\raggedright \small \\ \quad \quad \quad b. All analyses were controlled for designers' differences
     (gender and chatbot experience). Whenever applicable,
     analyses (e.g., \textit{informativeness}) were additionally
     controlled for respondents' differences (the corresponding metric
     in Q1).}
     {\raggedright \small \\  \noindent c. \textit{informativeness} and \textit{length} were additionally controlled for \textit{engagement duration}; and \textit{engagement duration} was controlled for \textit{length} \cite{Xiao2020-fr}. }

  \label{tb:4}
 \end{table*}
\subsubsection{\textbf{\textit{iChatProfile} helped designers build better chatbots (RQ1)}}
\smallskip

Table~\ref{tb:3} and Table~\ref{tb:4} summarize the analysis
results. The results show that \textit{iChatProfile} helped create
chatbots that performed significantly better at both interview and
individual question level. At the interview level, for example, the
chatbots in Group B achieved a higher \textit{completion rate} (78\%
vs. 74\% vs. 74\%) and more positive \textit{user sentiment} (66\%
vs. 45\% vs 46\%), than those in Group w/o  \textit{iChatProfile} and the baseline.

 At the question level, the chatbots in  Group w/ \textit{iChatProfile}  also performed better
 than those in Group w/o  \textit{iChatProfile} and the baseline on almost all dimensions,
 including \textbf{response quality} (\textit{informativeness}),
 \textbf{user engagement} (\textit{response length} and
 \textit{engagement duration}), and \textbf{user experience}
 (\textit{level of empathy} and \textit{repetition rate}). Only the
 difference in \textit{privacy intrusion rate} is insignificant. This
 is because the interview questions (Table~\ref{tb:questiontable}) did
 not elicit much private or sensitive user information. Our results
 also indicated that the performance differences between the chatbots
 in  Group w/o  \textit{iChatProfile} and the baseline are mostly insignificant. In fact, Group w/o  \textit{iChatProfile} performed \textit{worse} than the baseline on certain metrics,
such as \textit{completion rate} and \textit{user satisfaction
   rating}. This implies that designers had difficulty improving a
 chatbot without any specific design guidance. Moreover, making
   improvements  without knowing chatbot deficiencies could even hurt the chatbot performance. For example,
   a designer in Group w/o  \textit{iChatProfile} added a follow-up question \textit{"What made
     you feel that way”} to interview question Q4 \textit{"What
     challenges are you facing”}. But he ignored user responses that
   already talked about their feelings when answering Q4, which made
   users feel unheard. No wonder one user commented \textit{”I have
     already stated that, you were unable to understand answers.”}

 In all the analyses, the use of \textit {iChatProfile} was a
 significant factor impacting the chatbot performance
 differences. Three control variables, Q1, \textit{gender}, and
 \textit {chatbot experience}, were shown significant for a few
 analyses, although none of these control variables had interaction
 effect with the use of \textit{iChatProfile}. Control variable Q1 was
 significant for \textit{informativeness}, \textit{response length},
 and \textit{engagement duration}.  Since Q1 was used to account for
 respondents' differences \cite{Xiao2019-hk}, the effect of Q1 implies
 the effect of respondents' behavior (e.g., uncooperation) on
 interview quality, consistent with previous findings
 \cite{Xiao2019-hk}. In addition, control variable gender
 significantly impacted a chatbot's empathy level, and one's chatbot
 experience influenced \textit{informativeness} and \textit{response
   length}.  It is interesting that the chatbots made by male
 designers were more empathetic than those made by female
 designers ( Male 0.026 vs. Female 0.003, p<0.05 ). Although one's chatbot experience helped make chatbots
 better at eliciting information (e.g., \textit{informativeness}), it
 had no effect on user experience, such as \textit{user satisfaction
   rating} or \textit{user trust rating}.

\subsubsection{\textbf{\textit{iChatProfile} helped designers make chatbot design decisions (RQ2)}} During the post-task interviews, all participants from Group w/ \textit{iChatProfile} 
confirmed the helpfulness of \textit{iChatProfile} and benefited from
the displayed chatbot profile, design suggestions, and evidential
conversation examples. Specifically, a chatbot profile provided
designers with an overview of the chatbot performance. They liked the
visual display because it provided them with \textit{"straightforward
  insights"}. They also mentioned that they used the metric scores
shown in the chatbot profile (Fig~\ref{fig:3}(a)) to quickly locate
design deficiencies in the baseline chatbot.

\vspace{\baselineskip} 
\noindent \textit{Guiding Designers to Make Practical Chatbot Improvements}

\noindent Recall that at the beginning of their task, all designers
were asked to comment on the baseline chatbot and their plan to
improve it. All of them gave vague descriptions or improvement
plans. However, after the designers in  Group w/ \textit{iChatProfile}  had access to
\textit{iChatProfile}, they seemed knowing what they needed to do. On
average, the designers in  Group w/ \textit{iChatProfile}  did 495\% (99 vs. 20) of chatbot
customizations compared to those in  Group w/o  \textit{iChatProfile}  (Table~\ref{tb:designcustomization}).

Specifically, the designers in  Group w/ \textit{iChatProfile} appreciated the design
suggestions and evidential conversation examples. 
We  examined the chatbots made by the designers in  Group w/ \textit{iChatProfile}  and observed
that all of them followed one or more design suggestions given by
\textit{iChatProfile}. For example, the two design suggestions,
\textit{"add customizations to show the chatbot is actively
  listening"} and \textit{"reword the question to make it more
  acceptable to users"}, were followed by all 5 designers in  Group w/ \textit{iChatProfile} 
to customize and improve the baseline chatbot behavior around
interview questions Q2 and Q4 (Table~\ref{tb:questiontable}). In addition to following the design suggestions, the designers in
 Group w/ \textit{iChatProfile}  also found the conversation examples very helpful. While the
design guidelines informed designers \textit{what} to do (e.g.,
rewording a question), the conversation examples helped them figure
out \textit{how} to do it. For example, Q2 \textit{``where are you located''} asked
respondents about their location. When chatting with the baseline
chatbot, some respondents were unclear about the question, which
caused a higher \textit{repetition rate}. \textit{iChatProfile}
generated a design suggestion \textit{"reword the question to make it
  more acceptable to users"} with conversation examples (Table~\ref{tb:repetitionexample1}-\ref{tb:repetitionexample2}). One designer who followed the suggestion reworded the
original question to \textit{``May I ask where are you located? No
  need to be very specific, just city name would do :)''}. He stated
in the post-task interview that \textit{"I read the bad example and
  realized that some people might not like this kind of questions
  directly asking for their personal information, so I changed it."}

\begin{table*}[htb!]
\renewcommand\arraystretch{1.2}
\centering
\captionsetup{justification=centering}
\caption{A summary of chatbot customizations made by designers \\ in Group A (w/o iChatProfile) and Group B (w/ iChatProfile)}
\begin{tabular}{l!{\color{black}\vrule}ccc!{\color{black}\vrule}ccc} 

\multirow{2}{*}{~}  & \multicolumn{3}{c|}{Group A} & \multicolumn{3}{c}{Group B} \\ 
\cline{2-4}
\cline{5-7} & Mismatched    & Matched & Total count    & Mismatched & Matched    & Total count \\ 
\hline
Improving response quality  & 5 (55.6\%) & 4 (44.4\%)      & 9  & 10 (45.5\%) & 12 (54.5\%) & 22       \\ 
\hline
Improving user engagement  & 5 (71.4\%)  & 2 (28.6\%)  & 7   & 15 (42.9\%)  & 20 (57.1\%) & 35\\ 
\hline
Improving user experience  & 0 (0\%) & 4 (100\%) & 4   & 4 (9.5\%) & 38 (90.5\%)  & 42    \\
\hline
Improving  ethics  & 0  & 0  & 0 & 0  & 0  & \multicolumn{1}{c!{\color{white}\vrule}}{0}  \\\hline
\end{tabular}
\vspace{1ex}
    \small \\"Mismatched"/"Matched" indicate whether designers'
    chatbot customizations matched with what \textit{iChatProfile}
    suggested.

\label{tb:designcustomization}
\end{table*}

In comparison, the chatbot
customizations made by designers in  Group w/o  \textit{iChatProfile}  were fewer and with a
higher percentage of \textit{unmatched} chatbot customizations (43\%
vs. 33\% in  Group w/ \textit{iChatProfile} ) — customizations that were not suggested by
\textit{iChatProfile}. Table~\ref{tb:designcustomization} shows the design suggestions given
by \textit{iChatProfile} and the chatbot customizations made by
designers two groups, respectively. From our observations, none of the designers from Group w/o \textit{iChatProfile} went through all the 128 interviewee responses. They mainly reviewed the report dashboard and randomly selected a few responses to examine, without specifically knowing what to learn from the interview results let alone how to improve the chat based on the results.  Without any guidance from a tool like
\textit{iChatProfile}, the designers in  Group w/o  \textit{iChatProfile}  made their
customizations based on their intuition or ad hoc reasons. For
example, when asked \textit{"why did you decide to add these
  customizations"}, one designer in  Group w/o  \textit{iChatProfile}  said \textit{"Because the
  functionality of adding customization provided by Juji is a large
  block (a big area of the interface). It is very noticeable and I
  decided to try it out."}. Another designer also stated that \textit{"Juji provides so many features to choose from and I’m not sure which one to use ... I decide to add them all in the end."}. Since the designers in  Group w/o  \textit{iChatProfile}  didn't use \textit{iChatProfile}, a lack of guidance for identifying chatbot
deficiencies or improvements definitely contributed to their inferior chatbot performance.

\vspace{\baselineskip} 
\noindent \textit{Inspiring Designers to Make Creative Chatbot Improvements}
 
In addition to guiding designers to make practical chatbot
improvements, \textit{iChatProfile} also inspired designers to make
creative chatbot improvements beyond what was suggested by the
tool. For example, one designer in  Group w/ \textit{iChatProfile}  decided to add transitions
between interview questions, \textit{"Come on, get up and do 10 bobby
  jumps before we continue. Cheer up!"}. During the post-task
interview, when asked why he made such a design decision, he mentioned
that \textit{"I noticed the profile shows the original design
  (baseline) did not engage people well. So I thought why not engage
  them physically?"} We also checked the feedback left by the
respondents who chatted with this chatbot. We noticed the positive comments
such as \textit{"You (Juji) asked me to do bobby jumps. I didn't
  actually do it but it's interesting and I like it."}

\section{Discussions}

While our study results are encouraging, the study also revealed
several limitations. Here we discuss these limitations and future work. We also briefly discuss design implications of our
work on building chatbots beyond interview chatbots.

\subsection {Limitations}

\subsubsection {Study Scope and Participants}
While our results should be widely applicable for building
  a class of interview chatbots, the scope and the participants of our
  study present its limitations. Our study reported here focused on a
  low-stakes interview task (e.g., user or market
  research interviews) and recruited all the interviewees on Amazon Mechanic
  Turk. It is unclear whether our results would hold for different
  types of interview tasks, such as high-stakes tasks like job
  interviews with much more motivated interviewees and additional
  chatbot requirements (e.g., detecting faking
  \cite{zhou2019trusting}). Moreover, in our study all the chatbot
  designers were university students. It would be interesting to
  investigate how our results would hold or change with different
  chatbot designer groups (e.g., experienced chatbot designers).`

\subsubsection {Offering Finer-Grained Design Suggestions}
\smallskip

Currently, \textit{iChatProfile} often offers multiple design
suggestions per performance metric. For example, it offers two
suggestions if the \textit{response length} is below a threshold
(Table~\ref{tb:designguideline}). However, under certain circumstances, one suggestion might
be more useful than others. Using the example for improving
\textit{response quality} due to a vague question, adding explanations
to the question would be more useful than making the question more
acceptable to users. This requires that \textit{iChatProfile} further
discerns the \textit{causes} to the chatbot performance so it can
narrow down the design suggestions and recommend the most suitable
one. One potential method to address this is to analyze user responses
and identify different semantic themes, similar to the data-driven
methods used by others to recognize the semantic themes in user input
\cite{Xiao2020-fr}. Based on the recognized themes and the computed
performance metric, \textit{iChatProfile} can recommend the most suitable
design suggestion(s).

Additionally, \textit{iChatProfile} currently produces evidential
conversation examples along with design suggestions, which proved to
be helpful for designers in our study. However, these examples are the
``negative examples'' extracted from existing chats and no positive
examples are given. For example, if the design suggestion is to
\textit{``give empathetic feedback''}, it would be helpful for a
designer to see a ``positive example''---what an empathetic feedback
is like. Again, this would require more usage data, which will then allow
\textit{iChatProfile} to extract ``good examples''.

\subsubsection {Evaluating \textit{iChatProfile} Usability}
\smallskip 
Although our ultimate goal is to help designers build effective
interview chatbots, we have not yet evaluated the usability of
\textit{iChatProfile} for two reasons. First, we want to verify its
usefulness and effect before evaluating its usability. Second, our
current implementation is standalone and not integrated with any
chatbot platforms. Thus certain operations are cumbersome involving
much manual work (e.g., manually downloading all the chat transcripts
from a chatbot platform and then uploading them into
\textit{iChatProfile}). A more integrated version should be created and
then usability evaluation makes better sense.

\subsection {Future Work and Design Implications}
There are several directions that we can extend \textit{iChatProfile} to refine its functions and expand its uses. 

\subsubsection{System Explainability}
\smallskip

As mentioned in Section 6.3.1, \textit{iChatProfile} provides
explanations on \textit{why} certain design suggestions are given. Our
study participants expressed their appreciation of such explanations.
However, when system suggestions were inconsistent with designers'
belief, current explanations need to be expanded. For example, one
designer stated \textit{"The score of trust level doesn't actually reflect my
  experience, I am wondering why."}. In such cases, deeper
explanations would be helpful than just stating \textit{``the score is lower than a
  threshold\"} .  

One possible direction of future work is
to construct a multi-layer framework for evaluating the performance of
interview chatbots. Following \cite{Pu2011-bi}, this framework could
have three layers: \textbf{design quality} (e.g., the question-level
performance metrics used in our current work), \textbf{user belief}
(user perceived chatbot performance, such as perceived
usefulness\cite{pu2006trust} and ease of use
\cite{chen2009interaction}), and \textbf{user attitude} (users'
overall feelings towards the whole chatbot, such as perceived trust
and satisfaction in our work). A path model can then be generated to
reveal causal relationships between different layers to make chatbot
profiling and design suggestions more explainable. In particular, such
relationships could explicate how design qualities may influence
users' attitude more clearly, or why a certain metric could contribute
to the success/failure of the overall design through an influence path
across layers \cite{Pu2011-bi}.

\subsubsection {Benchmarking Interview Chatbot Evaluation}

\smallskip

As our studies show, it is difficult for designers to
evaluate chatbot performance without any
guidance. While \textit{iChatProfile} helps designers make specific chatbot improvements,
it could not inform the designers \textit{how much} test an interview chatbot
actually needs before achieving an acceptable performance. For
example, it would be very valuable to inform a chatbot designer that a
minimal of \textit{N} pilot users or at least \textit{M} rounds of
evaluations are needed to test a particular interview chatbot and
achieve an acceptable performance. As \textit{iChatProfile} collects
more usage data, offering designers with the above suggestions will
become more feasible. Specifically, an evaluation benchmark can be
established for each type of interview question as well as for a
particular type of interview chatbot. This will also enable us to
establish different thresholds for different chatbot tasks.

\subsubsection{Real-Time Chatbot Evaluation and Feedback}
\smallskip

Although we showed how \textit{iChatProfile} was able to
  help improve chatbot performance significantly within just one
  iteration of design, building an effective interview chatbot often
  takes multiple iterations.  Currently, \textit{iChatProfile}
generates a chatbot profile and design suggestions \textit{after} a
designer deploys the chatbot and collects a set of transcripts from
live chats. However, designers may wish to receive prompt feedback
while designing a chatbot, since early and timely feedback could
improve creative work\cite{kulkarni2014early,
  kulkarni2015peerstudio}. To enable continuous chatbot
  evaluation and improvement, one approach is to integrate
  \textit{iChatProfile} with crowdscourcing tools that can recruit
  testers, administer live chats, and provide a chatbot profile and
  design suggestions to improve the chatbot, all near real time
  \cite{Mitchell_Gordon_University_of_Rochester_Rochester_NY_USA_undated-eq}.
  Another approach is to employ deep learning algorithms to simulate
  real-world user behaviors so that chat transcripts can be obtained
  in real time \cite{Deriu2019-wo} to aid the iterative evaluation and
  improvement of chatbots in real time.
  
\subsubsection{Assistive Design of Chatbots beyond Interview Chatbots} Our work demonstrates the effectiveness of \textit{iChatProfile} for
helping designers to evaluate and improve an interview chatbot
iteratively. Since interview chatbots share many characteristics as
other types of chatbots, such as counseling or training chatbots,
\textit{iChatProfile} could be extended to help designers build such
chatbots as well. Especially \textit{iChatProfile} aims at helping non-AI
experts in chatbot design, it could help professional coaches or
trainers design, evaluate, and improve their own chatbots. We hope
that our work can serve as a stepping-stone on the path to
democratize chatbot design for a wide variety of applications beyond
interview tasks.

\section{Conclusions}

We described a computational framework for evaluating interview chatbots
and presented \textit{iChatProfile}, a tool that
helps  designers to evaluate and improve interview chatbots
iteratively. Given a set of chat transcripts, it automatically
quantifies the performance of a chatbot and generates a chatbot
profile. Based on the generated chatbot profile, it also offers design
suggestions in natural language with evidential conversation examples,
which help guide designers to improve the chatbot. To validate the
effectiveness of \textit{iChatProfile}, we designed and conducted a
between-subject study that compared the performance of chatbots
designed with and without using \textit{iChatProfile}. Based on the
transcripts collected from the live chats between 10 chatbots and 1394 users,
our results show that \textit{iChatProfile} helped produce interview
chatbots with significantly better performance across almost all
dimensions, including response quality, user engagement, and user
experience.

\bibliographystyle{ACM-Reference-Format}
\bibliography{acmart}

\end{document}